\documentclass[aps,floatfix,preprint,nofootinbib]{revtex4-1}
\usepackage[parfill]{parskip}    
\usepackage{graphicx}
\usepackage{amssymb}
\usepackage{epstopdf}
\usepackage{verbatim}
\usepackage{hyperref}
\usepackage{url}
\usepackage{setspace}
\newcommand{\be}{\begin{equation}}
\newcommand{\bea}{\begin{eqnarray}}
\newcommand{\beq}[1]{\begin{equation}\label{#1}}

\newcommand{\ee}{\end{equation}}
\newcommand{\eea}{\end{eqnarray}}
\newcommand{\eeq}{\end{equation}}
\newcommand{\lsim}{\!\mathrel{\hbox{\rlap{\lower.55ex \hbox{$\sim$}} \kern-.34em \raise.4ex \hbox{$<$}}}}
\newcommand{\gsim}{\!\mathrel{\hbox{\rlap{\lower.55ex \hbox{$\sim$}} \kern-.34em \raise.4ex \hbox{$>$}}}}

\newcommand{\abs}[1]{\left| #1 \right|}

\begin{document}

\setlength{\baselineskip}{0.22in}

\begin{flushright}MCTP-12-02
\end{flushright}
\vspace{0.2cm}

\title{Neutrinos from Off-Shell Final States and the Indirect Detection of Dark Matter}
\author{John Kearney and Aaron Pierce}
\vspace{0.2cm}
\affiliation{Michigan Center for Theoretical Physics (MCTP) \\
Department of Physics, University of Michigan, Ann Arbor, MI
48109}

\date{\today}

\begin{abstract}
We revisit the annihilation of dark matter to neutrinos in the Sun near the $W^+ W^-$ and $t \bar{t}$ kinematic thresholds.  
We investigate the potential importance of annihilation to $WW^\ast$ in a minimal dark matter model in which a Majorana singlet is mixed with a vector-like electroweak doublet, but many results generalize to other models of weakly-interacting dark matter.  We re-evaluate the indirect detection constraints on this model and find that, once all annihilation channels are properly taken into account, the most stringent constraints on spin-dependent scattering for dark matter mass $60 \text{ GeV} \lsim m_\chi \lsim m_t$ are derived from the results of the Super-Kamiokande experiment.
Moreover, we establish the model-independent statement that Majorana dark matter whose thermal relic abundance and neutrino signals are both controlled by annihilation via an $s$-channel $Z$ boson is excluded for $70 \text{ GeV} \lsim m_\chi \lsim m_W$.  In some models, annihilation to $tt^\ast$ can affect indirect detection, notably by competing with annihilation to gauge boson final states and thereby weakening neutrino signals.  However, in the minimal model, this final state is largely negligible, only allowing dark matter with mass a few GeV below the top quark mass to evade exclusion.
\end{abstract}

\maketitle

\section{Introduction}\label{sec:intro}

The accumulation and subsequent annihilation of dark matter in the Sun could lead to a significant flux of high-energy neutrinos discernible from background \cite{Press:1985ug,Silk:1985ax, Srednicki:1986vj, Jungman:1995df}.  Alternately, a lack of signal can be used to place limits on the rate of solar dark matter annihilation.  Because capture and annihilation are often in equilibrium,  this approach typically probes the dark matter capture rate, or equivalently dark matter-nucleon scattering cross sections.  As such, it gives information on some of the same couplings probed in underground direct detection experiments.  Generally, constraints on spin-independent (SI) scattering from indirect detection of neutrinos are substantially weaker than those  
from direct detection experiments.  However, in sizable portions of parameter space, indirect detection provides the most stringent constraints on spin-dependent (SD) scattering \cite{Kamionkowski:1994dp, Wikstrom:2009kw}.  The promise of additional data from DeepCore \cite{DeYoung:2011ke, Ha:2012ww} motivates revisiting
 the calculation of neutrino fluxes.

Studies of neutrino spectra and signals from dark matter annihilations generally concentrate on 2-body final states -- see, for instance, \cite{Cirelli:2005gh, Blennow:2007tw}.  However, there are cases in which  3-body final states may contribute significantly to neutrino signals in spite of the 
phase space suppression they suffer relative to 2-body final states.  For example, if the dark matter mass is just below the energy required to open up annihilation to a new 2-body final state, the rate of annihilation to a corresponding 3-body final state can be sizable.  This has been studied previously in the Minimal Supersymmetric Standard Model (MSSM) for dark matter annihilation to $WW^\ast$ and $tt^\ast$ \cite{Chen:1998pp}.  Another example arises wherein spin-1 electroweak boson emission is capable of lifting helicity suppression in certain $2 \rightarrow 2$ processes, which can lead to sizable branching ratios for dark matter annihilation to 3-body final states consisting of the two original final-state particles and an additional spin-1 boson \cite{Bergstrom:1989jr,Bringmann:2007nk,Ciafaloni:2010ti,Ciafaloni:2011sa}.  Furthermore, the neutrinos produced by these 3-body final states can be more energetic than those from the dominant 2-body channels, enhancing the importance of 3-body final states.

%

In this paper, we revisit the importance of annihilation to 3-body states just below 2-body thresholds for weakly-interacting dark matter.  We build upon and generalize some results from \cite{Chen:1998pp}.   In section~\ref{sec:wwast}, we discuss dark matter annihilation to $WW^\ast$.  As a  
representative candidate for
weakly-interacting dark matter, we 
use the minimal model studied in \cite{ArkaniHamed:2005yv,Mahbubani:2005pt, Enberg:2007rp, D'Eramo:2007ga,Cohen:2011ec}, see also \cite{Fayet:1974fj}.  In this model, the dark matter is a Majorana fermion composed of an admixture of a weak doublet and a sterile singlet.  As such, it is similar to a mixed Bino--Higgsinso state of the MSSM.  
Not only is the ``singlet-doublet" model an interesting dark matter candidate in its own right, but results we obtain for dark matter annihilating to $WW^\ast$ are also applicable to a range of weakly-interacting dark matter models.  These results are presented in section~\ref{sec:wwastgen}.  To highlight the potential significance of 3-body final states, we assess the overall importance of annihilation to $WW^\ast$ in the minimal model in section~\ref{sec:wwastsdm}.  In section~\ref{sec:ttast}, we turn our attention to $tt^\ast$, presenting general results for dark matter annihilation to $tt^\ast$ via an $s$-channel vector boson.  Motivated by our findings in section~\ref{sec:wwastsdm}, we re-evaluate the indirect detection limits on the singlet-doublet model in section~\ref{sec:revisedsdm}, including the effects of subdominant annihilation channels.  The revised limits are significantly stronger than those originally quoted by these authors in \cite{Cohen:2011ec}.  
Conclusions are presented in section~\ref{sec:conclusion}.

\section{$WW^\ast$}\label{sec:wwast}

Electroweakly-interacting dark matter with mass $m_\chi < m_W$ will annihilate to 2-body final states consisting of pairs of Standard Model fermions, $\chi \chi \rightarrow f \bar{f}$.
The dark matter may also annihilate to final states of the form $W f \bar{f}'$, where $f, \bar{f}'$ denote Standard Model fermions that arise from an off-shell $W^\ast$.  
Near threshold, 
annihilation to $WW^\ast$ 
may contribute 
to neutrino signals.  

This is particularly true in the case of Majorana dark matter for two reasons.  First, dark matter particles in the Sun have velocities $v \sim 10^{-4} - 10^{-3}$, so the $v \rightarrow 0$ static limit 
applies.  In this limit, only the $s$-wave component of cross sections survives; for annihilation to fermion pairs, helicity arguments require a suppression of $(m_f/m_Z)^2$. 
Second, the mass dependence favors the $\chi \chi \rightarrow b \bar{b}$ process (with subdominant contributions from $\tau^+ \tau^-$ and $c \bar{c}$).  Neutrinos from $W$'s tend to be significantly harder than those from $b$'s, and the presence of additional, particularly energetic neutrinos from the $W f \bar{f}^\prime$ final state could enhance neutrino signals.



The singlet-doublet model consists of a gauge singlet fermion and a pair of fermionic $SU(2)$ doublets.  The doublets have a vector-like mass term, and the neutral components of the doublets mix with the gauge singlet through renormalizable couplings to the Higgs field, $H$. The new fields are odd under a $\mathbb{Z}_2$ symmetry, ensuring the stability of the lightest state.  We denote the singlet as $N$ and the doublets as $D, D^c$:
\begin{eqnarray}
D = \left(\begin{array}{c} \nu \\ E \end{array} \right) & \qquad & D^c = \left(\begin{array}{c} -E^c \\ \nu^c \end{array}\right),
\end{eqnarray}
with hypercharges $-\frac{1}{2}$ and $+\frac{1}{2}$ respectively.  Mass terms and interactions for this model are given by:
\begin{equation}
\Delta \mathcal{L} = - \lambda D H N - \lambda' \tilde{H} D^c N - m_D D D^c - \frac{1}{2} m_N N^2 + \mbox{ h.c.},
\end{equation}
where $SU(2)$ doublets are contracted with the Levi-Civita symbol $\epsilon^{ij}$ and $\tilde{H} \equiv i \sigma_2 H^\ast$.  When the Higgs field attains its vacuum expectation value, the singlet $N$ and the neutral components $\nu$ and $\nu^c$ of the doublets mix, such that the spectrum consists of a charged Dirac fermion of mass $m_D$, which we denote $E^\pm$, and three neutral Majorana fermions ($\nu_i$, $i = 1,2,3$), the lightest of which is the dark matter.  Details can be found in \cite{Cohen:2011ec}.  
This dark matter candidate couples only to the bosons of the electroweak theory, and not additional exotic states, so in this sense is minimal.  The singlet-doublet mixing 
produces
Majorana dark matter, evading the much too large $Z$-mediated spin-independent cross sections that Dirac dark matter exhibits.  We view this model as a useful bellwether -- it provides an indication of the status of very simple dark matter models without special features (such as stau or stop co-annihilation in the MSSM).

We implemented the singlet-doublet model in MadGraph \cite{Alwall:2007st}, and simulated dark matter annihilations at $\sqrt{s}$ 
corresponding to $v \sim 10^{-3}$,  approximately reproducing the conditions of dark matter annihilations in the Sun.
We then decayed unstable particles using the MadGraph DECAY package.  
The unweighted event output was modified\footnote{ID codes for the incoming dark matter particles were changed to those corresponding to $e^+ e^-$ annihilation.} such that it could be passed to \textsc{Pythia} for showering and hadronization \cite{Sjostrand:2006za}.
Relevant data about neutrinos and their parents was extracted for each event, and fed to a modified version of WimpSim in order to simulate neutrino interaction and propagation to a detector (either IceCube/DeepCore or Super-K) \cite{wimpsimcode}.\footnote{For propagation, we use the default WimpSim parameters: $\theta_{12} = 33.2^\circ$, $\theta_{13} = 0.0^\circ$, $\theta_{23} = 45.0^\circ$, $\delta = 0.0$, $\abs{\Delta m_{21}}^2 = 8.1 \times 10^{-5} \text{ eV}^2$ and $\abs{\Delta m_{31}}^2 = 2.2 \times 10^{-3} \text{ eV}^2$ \cite{Maltoni:2004ei}, but our results are not particularly sensitive to the exact choice of these parameters.}
To validate this method,  we confirmed that the cross sections and branching ratios given by MadGraph agreed with those from analytic expressions for $2 \rightarrow 2$ annihilations in the static limit (e.g., from \cite{Dreiner:2008tw}).  Furthermore, we confirmed that the spectra given for $2 \rightarrow 2$ annihilations were the same as those given by the unmodified version of WimpSim and \cite{Cirelli:2005gh}.  

\subsection{General Results for $WW^\ast$ Neutrino and Muon Spectra}\label{sec:wwastgen}

\begin{figure}
\begin{center}
\includegraphics[width=0.75\textwidth]{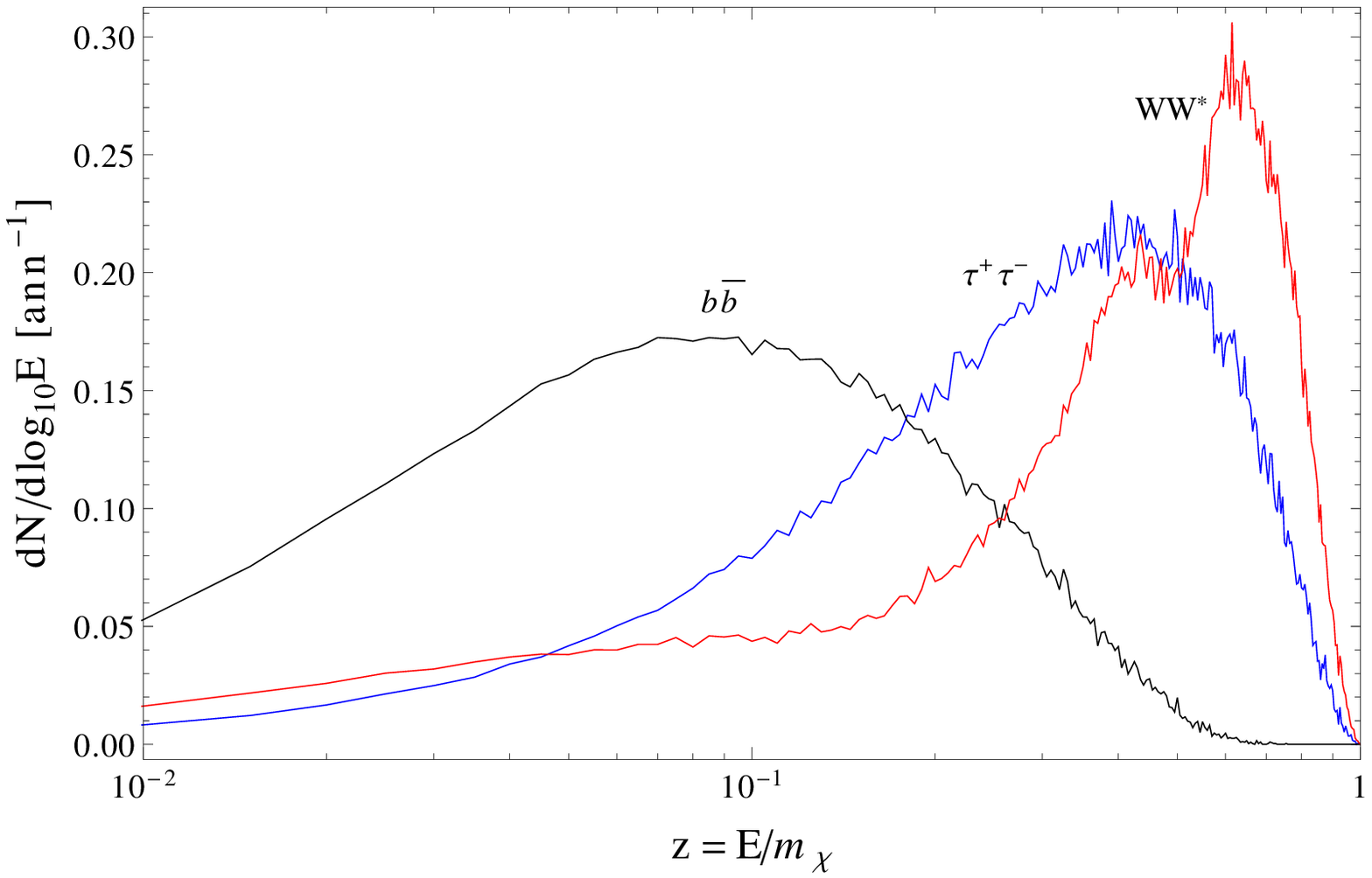}
\includegraphics[width=0.75\textwidth]{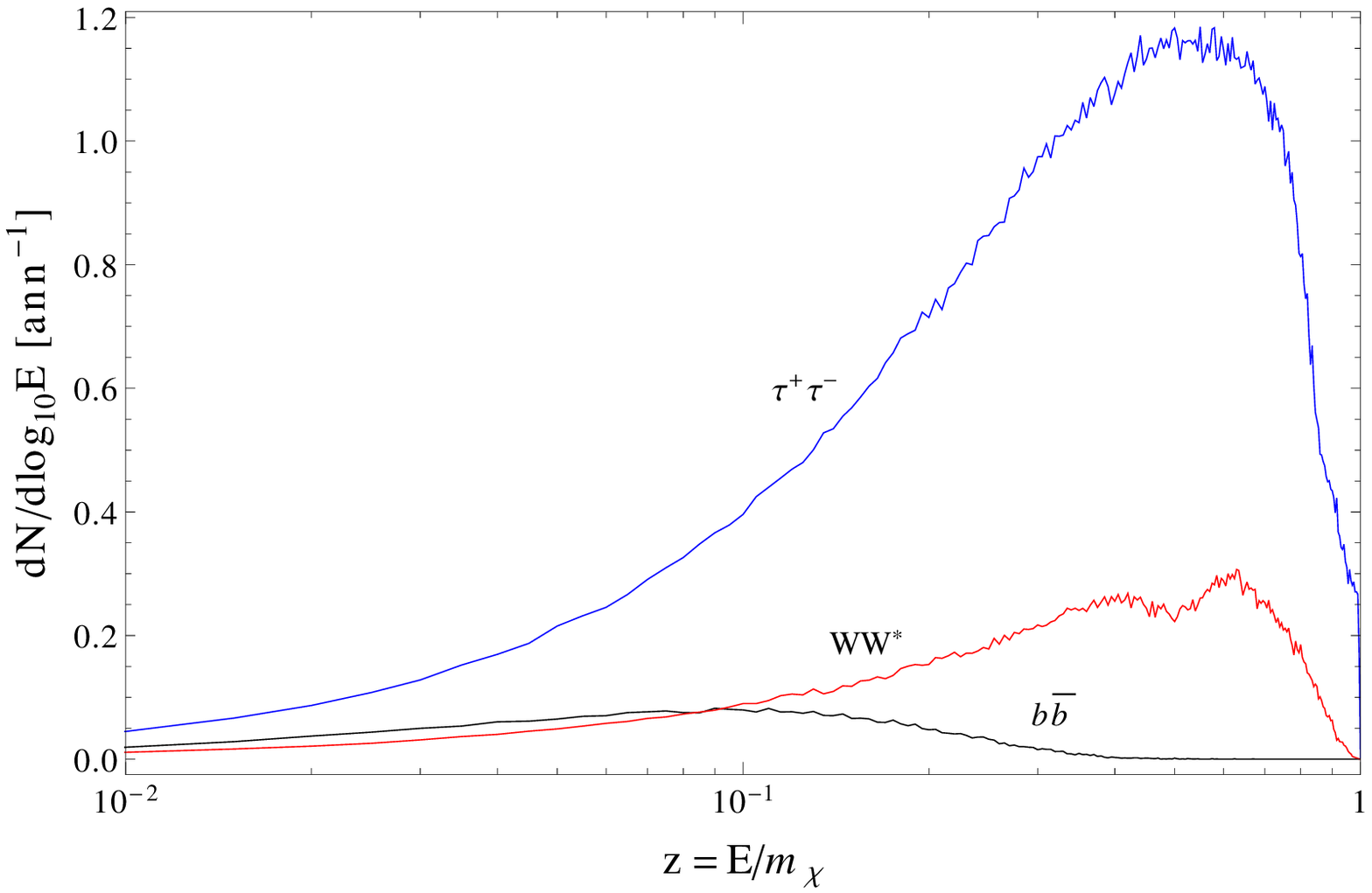}
\end{center}
\caption{\label{fig:WWastinjection}Neutrino injection spectra from annihilation to $b \bar{b}$ (black), $\tau^+ \tau^-$ (blue) and $W W^\ast$ (red) for dark matter mass $m_\chi = 75 \text{ GeV}$.  Shown are the spectra for $\nu_e$ (top) and $\nu_\tau$ (bottom) -- note that $\nu_e = \bar{\nu}_e = \nu_\mu = \bar{\nu}_\mu$ and $\nu_\tau = \bar{\nu}_\tau$.}
\end{figure}

\begin{figure}
\begin{center}
\includegraphics[width=0.75\textwidth]{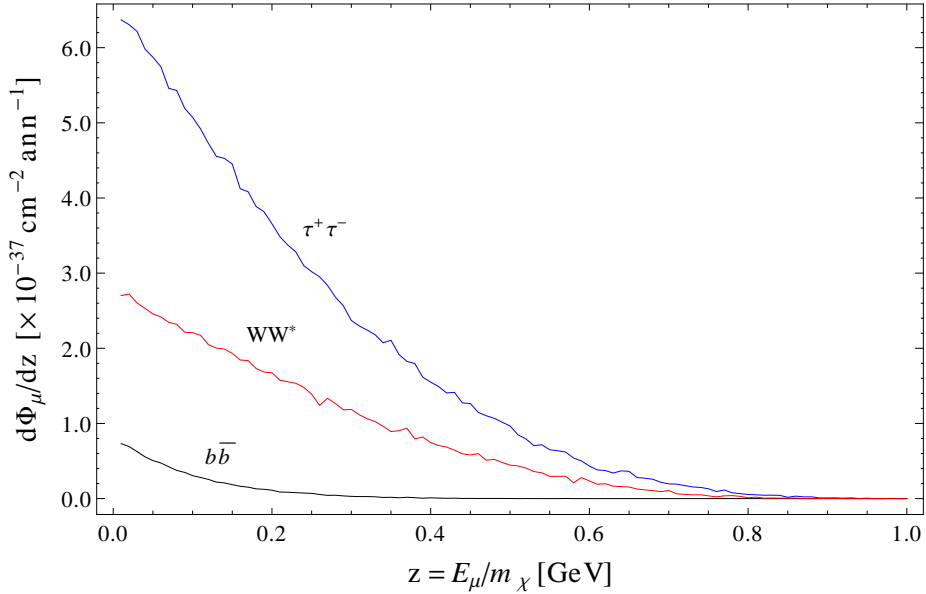}
\end{center}
\caption{\label{fig:WWmuons}Spectrum of muons per annihilation at DeepCore/IceCube from annihilation to $b \bar{b}$ (black), $\tau^+ \tau^-$ (blue) and $W W^\ast$ (red) for dark matter mass $m_\chi = 75 \text{ GeV}$.  Shown are the fluxes given by WimpSim at a plane at the location of the detector.  We emphasize that the curves shown correspond to a single annihilation to a given final state -- the (model-dependent) branching ratios to the different states have not been taken into account here.}
\end{figure}

Injection spectra for dark matter annihilation to $b \bar{b}$, $\tau^+ \tau^-$ and $W W^\ast$ based on a simulation of $10^6$ events are shown in figure~\ref{fig:WWastinjection} for $m_\chi = 75 \text{ GeV}$.  The corresponding spectra of muons at DeepCore for each annihilation channel is shown in figure~\ref{fig:WWmuons}.  As anticipated, the neutrino and muon spectra from annihilation to $WW^\ast$ are significantly harder than those from annihilation to $b \bar{b}$, although they are softer than those from $\tau^+ \tau^-$.  Recall, $\tau$ decays produce energetic $\nu_\tau$ that can oscillate to give $\nu_\mu$ at a detector.  

Integration gives the total flux of muons per annihilation above a threshold energy $E_\mu^{\text{thresh}}$, $\Phi_\mu^{\text{final state}}(E_\mu \ge E_\mu^{\text{thresh}})$, which can be used to determine the relative signal at a detector from each final state.  The fluxes for two different threshold energies, $E_\mu^{\text{thresh}} = 10 \text{ GeV}$ (projected for DeepCore \cite{DeYoung:2011ke}) and $E_\mu^{\text{thresh}} = 35 \text{ GeV}$ (a more conservative value quoted in \cite{Barger:2011em}), are given in table~\ref{tab:wwastflux}.  For $E_\mu^{\text{thresh}} = 10 \text{ GeV}$, the ratios of flux from $W f \bar{f}'$ to the fluxes from the 2-body final states (again, normalized to a single annihilation of each type) are
\begin{equation}
\frac{\Phi^{W f \bar{f}'}_\mu}{\Phi^{\tau^+ \tau^-}_\mu} (E_\mu \ge 10 \text{ GeV}) = 0.46, \qquad \frac{\Phi^{W f \bar{f}'}_\mu}{\Phi^{b \bar{b}}_\mu} (E_\mu \ge 10 \text{ GeV}) = 25.
\end{equation}
If we increase the threshold to $E_\mu^{\text{thresh}} = 35 \text{ GeV}$, the ratios become
\begin{equation}
\frac{\Phi^{W f \bar{f}'}_\mu}{\Phi^{\tau^+ \tau^-}_\mu} (E_\mu \ge 35 \text{ GeV}) = 0.46, \qquad \frac{\Phi^{W f \bar{f}'}_\mu}{\Phi^{b \bar{b}}_\mu} (E_\mu \ge 35 \text{ GeV}) = 670. 
\end{equation}
These values are fairly constant over the mass range $65 \text{ GeV} \leq m_\chi \leq m_W$.\footnote{Fixing $E_\mu^{\text{thresh}} = 10 \text{ GeV}$, for $m_\chi = 65 \text{ GeV}$ fluxes are approximately 0.8 times those given, and for $m_\chi = 80 \text{ GeV}$ about 1.2 times those given, with ratios remaining largely constant.}   The flux from annihilation to $W W^\ast$ is substantially larger than that from annihilation to $b \bar{b}$, and is comparable to that from annihilation to $\tau^+ \tau^-$.
Thus, if the branching ratio is not too small, $W f \bar{f}^\prime$ can conceivably contribute significantly to indirect detection signals.  For reference, muon fluxes at Super-K ($E^{\text{thresh}}_\mu \approx 2 \text{ GeV}$) are also given in the table.




\begin{table}
\setstretch{1.5}
\begin{center}
\begin{tabular}{| @{\hspace{0.2cm}} c @{\hspace{0.2cm}} | @{\hspace{0.2cm}} c @{\hspace{0.2cm}} | @{\hspace{0.2cm}} c @{\hspace{0.2cm}} || @{\hspace{0.2cm}} c @{\hspace{0.2cm}} |}
\hline
Final State & $\Phi_\mu^{\text{final state}} (E_\mu \ge 10 \text{ GeV})$ & $\Phi_\mu^{\text{final state}} (E_\mu \ge 35 \text{ GeV})$ &  $\Phi_\mu^{\text{final state}} (E_\mu \ge 2\text{ GeV})$\\ \hline
$b \bar{b}$ & $1.9 \times 10^{-39}$ & $1.0 \times 10^{-41}$  & $7.4 \times 10^{-39}$\\
$\tau^+ \tau^-$ & $1.0 \times 10^{-37}$ & $1.5 \times 10^{-38}$ & $1.9 \times 10^{-37}$\\
$W f \bar{f}^\prime$ & $4.6 \times 10^{-38}$ & $7.0 \times 10^{-39}$ & $8.5 \times 10^{-38}$ \\
\hline
\end{tabular}
\end{center}
\setstretch{1}
\caption{\label{tab:wwastflux}Fluxes of muons $[\text{cm}^{-2} \text{ ann}^{-1}]$ with energy $E_\mu \ge E_\mu^{\text{thresh}}$ at DeepCore/IceCube (first two columns) and Super-K (third column) from annihilations of dark matter with $m_\chi = 75 \text{ GeV}$ to various final states.}
\end{table}

Strictly-speaking, these values are model-dependent, as they depend on the polarization of the final state $W$'s.  
For singlet-doublet dark matter in the static limit, annihilation to $W f \bar{f}^\prime$ occurs via $t$- and $u$-channel exchange of the $E^\pm$.\footnote{The sum of the contributions to the $W f \bar{f}'$ final state from $s$-channel $Z$ exchange -- both where the $Z$ couples to $W W^\ast$ and where the $Z$ couples to a pair of fermions, one of which radiates a $W$ -- is small.}
Annihilation to $W$'s via exchange of a fermion in the $t$- or $u$-channel produces only transversely-polarized $W$ bosons (with the two polarizations in equal proportions).  In many models of electroweak dark matter (particularly in the $v \rightarrow$ 0 limit), it is precisely via $t$- and $u$-channel fermion exchange that annihilation to $W$ bosons occurs, so our results are certainly valid for these models.
But, in addition, the neutrino spectrum produced by annihilation to equal proportions of the three $W$ polarizations is similar to that from annihilation to equal proportions of the two transverse polarizations, and the differences between the overall results obtained from each are small.  
Thus, the values given in table~\ref{tab:wwastflux} 
are applicable to a range of models in which $WW^\ast$ states may be important and can be used to calculate the magnitude of the effect of these states in a given model -- all that must be known is the relevant annihilation branching ratios.

\subsection{$WW^\ast$ in the Singlet-Doublet Model}\label{sec:wwastsdm}


As a concrete example, we evaluate the importance of annihilation to $WW^\ast$ in the singlet-doublet model.  A sample point for which annihilation to $WW^\ast$ is important has $m_N = 75 \text{ GeV}$ and $m_D = 120 \text{ GeV}$, $\lambda = 0.384$ and $\lambda^\prime = -0.135$.  The values of $\lambda$ and $\lambda^{\prime}$ are fixed via the following reasoning.  

In \cite{Cohen:2011ec}, it was shown that for a light Higgs boson ($m_h \lsim 140 \text{ GeV}$), the dark matter-Higgs boson coupling must be relatively small to avoid SI direct detection constraints.  There exists a value \cite{Cohen:2011ec} where  the dark matter-Higgs boson coupling cancels completely, a value which we define as $\lambda^{\prime} \equiv \lambda^\prime_{\text{critical}}$.    While we choose $\lambda^{\prime} = \lambda^\prime_{\text{critical}}$, this choice of $\lambda^{\prime}$ is for simplicity; 
if the dark matter-Higgs boson coupling is sufficiently small to avoid direct detection constraints, it is not important for setting the relic density.\footnote{An exception occurs if $m_\chi \lsim \frac{m_h}{2}$ -- the enhancement to annihilation due to a small $s$-channel Higgs boson propagator allows the correct relic density to be achieved with a small dark matter-Higgs boson coupling, which generates spin-independent cross sections below experimental bounds.}
Moreover, dark matter annihilation through a scalar vanishes in the static limit, so a non-zero dark matter-Higgs boson coupling would not affect neutrino signals.\footnote{This also requires the spin-independent solar capture rate to be negligible for these points, which it is.} 
 We then fix the overall size of $\lambda$ by requiring the relic density $\Omega h^2 = 0.112$ (calculated using \texttt{micrOmegas} \cite{Belanger:2010gh}), consistent with the  determination by the seven-year Wilkinson Microwave Anisotropy Probe (WMAP) and other data on large scale structure \cite{Jarosik:2010iu}.  Note, if  $\lambda^\prime = \lambda^\prime_{\text{critical}}$, gauge invariance sets $m_\chi = m_N$.

For this choice of parameters, the correct relic density is achieved by $p$-wave annihilation via an $s$-channel $Z$ boson to Standard Model fermions in the early universe.  The annihilation 
is largely democratic amongst light fermions, taking advantage of the relatively large velocities ($v \sim \frac{1}{2}$) to avoid the significant $(m_f/m_Z)^2$ $s$-wave suppression.  Today in the Sun, dark matter velocities are sufficiently small that $p$-wave annihilation is negligible, and so annihilation occurs dominantly to $b \bar{b}$.  
This process is mass suppressed, permitting a non-trivial branching ratio for annihilation to $W f \bar{f}'$.
Branching ratios for the dominant solar annihilation channels are shown in table~\ref{tab:brwwast}, for both the case where only 2-body final states are considered and for the case where the $W f \bar{f}'$ final state is included.  While we give the branching ratio to $c \bar{c}$, 
its contribution to the muon flux is negligible.
These branching ratios can be used in conjunction with values of $\Phi_\mu^{\text{final state}}(E_\mu \ge E_\mu^{\text{thresh}})$ from the previous subsection to determine the model-specific indirect detection limits with and without annihilation to $WW^\ast$ for the sample point. 


\begin{table}
\setstretch{1.5}
\begin{center}
\begin{tabular}{| @{\hspace{0.2cm}} c @{\hspace{0.2cm}} | @{\hspace{0.2cm}} c @{\hspace{0.2cm}} c @{\hspace{0.2cm}} c @{\hspace{0.2cm}} c @{\hspace{0.2cm}} |}
\hline
 & $b \bar{b}$ & $c \bar{c}$ & $\tau^+ \tau^-$ & $W f \bar{f}'$ \\ \hline
2-body only & 86.9\% & 7.9\% & 5.1\% & -- \\
Including 3-body & 76.9\% & 7.0\% & 4.5\% & 11.5\% \\ \hline
\end{tabular}
\end{center}
\setstretch{1}
\caption{\label{tab:brwwast}Dominant branching ratios in the static limit for representative singlet-doublet model point with $m_\chi = m_N = 75 \text{ GeV}$, $m_D = 120 \text{ GeV}$, $\lambda = 0.384$ and $\lambda^\prime = \lambda^\prime_{\text{critical}} = -0.135$.  This point has $\Omega h^2 = 0.112$.  
The 2-body branching ratios are determined entirely by light fermion masses and color factors.  As a result, they apply in general for $m_b \ll m_\chi < m_W$.}
\end{table}
 

Recent results from Super-K give a model-independent limit on the total flux from dark matter annihilations of $\Phi_\mu(E_\mu \ge E_\mu^{\text{thresh}}) \leq 7.0 \times 10^{-15} \text{ cm}^{-2} \text{s}^{-1}$ for $m_\chi = 75 \text{ GeV}$ \cite{Tanaka:2011uf}.  This can be converted to a limit on $\sigma_{\text{SD}}^p$ by assuming the dark matter annihilates to a particular final state -- generally, ``soft'' limits are given by assuming annihilation to $b \bar{b}$.  Using the conversion factor of
\begin{equation}
\sigma_{\text{SD}}^p/\Phi_\mu = 7.6 \times 10^{11} \text{ cm}^2 \text{ s} \text{ pb}
\end{equation}
for annihilation exclusively to $b \bar{b}$ given in \cite{Wikstrom:2009kw}, the ``soft'' bound from Super-K is $\sigma_{\text{SD}}^p < 5.3 \times 10^{-3} \text{ pb}$ (as given in \cite{Tanaka:2011uf}).

This bound can be adapted to get the appropriate, model-dependent bound for the sample point by rescaling by the ratio of the average flux per dark matter annihilation for the sample point to the flux per annihilation exclusively to $b \bar{b}$.  Using the branching ratios from table~\ref{tab:brwwast} and the fluxes from table~\ref{tab:wwastflux}, we find these ratios to be
\begin{equation}
\frac{\Phi_\mu^{\text{ave, 2-body only}}}{\Phi_\mu^{b \bar{b}}} (E_\mu \ge 2 \text{ GeV}) = 2.2, \qquad \frac{\Phi_\mu^{\text{ave, incl. 3-body}}}{\Phi_\mu^{b \bar{b}}} (E_\mu \ge 2 \text{ GeV}) = 3.2.
\end{equation}
The numerators incorporate the flux from all relevant final states, appropriately weighted by branching ratios.
Thus, the model-dependent bounds for the sample point are
\begin{equation}
\label{eq:revlim}
\sigma_{\text{SD}}^{p,\text{2-body only}} < 2.4 \times 10^{-3} \text{ pb}, \qquad \sigma_{\text{SD}}^{p,\text{incl. 3-body}} < 1.6 \times 10^{-3} \text{ pb}.
\end{equation}
So including the effects of the $W f \bar{f}^\prime$ final state improves the bounds by a factor of
\begin{equation}
\frac{\Phi_\mu^{\text{ave, incl. 3-body}}}{\Phi_\mu^{\text{ave, 2-body only}}} (E_\mu \ge 2 \text{ GeV}) = 1.5.
\end{equation}

We note that both of these bounds are stronger than the bound of $\sigma_{\text{SD}}^p < 5.3 \times 10^{-3} \text{ pb}$ given in \cite{Tanaka:2011uf}.  This arises in part from appreciating the importance of the subdominant but hard $\tau^+ \tau^-$ channel.  Using standard assumptions, we find that these revised Super-K limits exclude the spin-dependent cross section for this particular point, $\sigma_{\text{SD}}^p = 3.2 \times 10^{-3} \text{ pb}$.  We can express the limits as what the local density of dark matter would have to be for this point to not be excluded.  We find the values to be
\begin{equation}
\rho^{\text{2-body only}} < 0.23 \text{ GeV cm}^{-3}, \qquad \rho^{\text{incl. 3-body}} <  0.15 \text{ GeV cm}^{-3}.
\end{equation}
Thus, this point is very tightly constrained by indirect detection experiments, if not excluded.  We quote the limits in terms of $\rho$, with the understanding that this is a placeholder for other astrophysical uncertainties.  For instance, the bound on $\sigma_{\text{SD}}^p$ goes as $\rho/\bar{v}$, where $\bar{v}$ represents the local dark matter velocity dispersion \cite{Jungman:1995df}.  So, while a decrease in $\rho$ could weaken the bound sufficiently for the sample point to avoid exclusion, so too could an increase in $\bar{v}$.  Recent analyses suggest that $\rho$ lies somewhere in the interval $[0.2,0.4] \text{ GeV cm}^{-3}$ \cite{Weber:2009pt, Catena:2009mf, Widrow:2008yg} and $\bar{v}$ may vary by up to $\mathcal{O}(20\%)$ \cite{McMillan:2009yr,Bovy:2009dr,Reid:2009nj}.\footnote{For a thorough, up-to-date review of these uncertainties, see \cite{Green:2011bv}.}  The effect of $W W^{\ast}$ on this point is sufficient to push it to a regime where there is substantial tension, even taking these uncertainties into account.

\begin{figure}
\begin{center}
\includegraphics[width=0.75\textwidth]{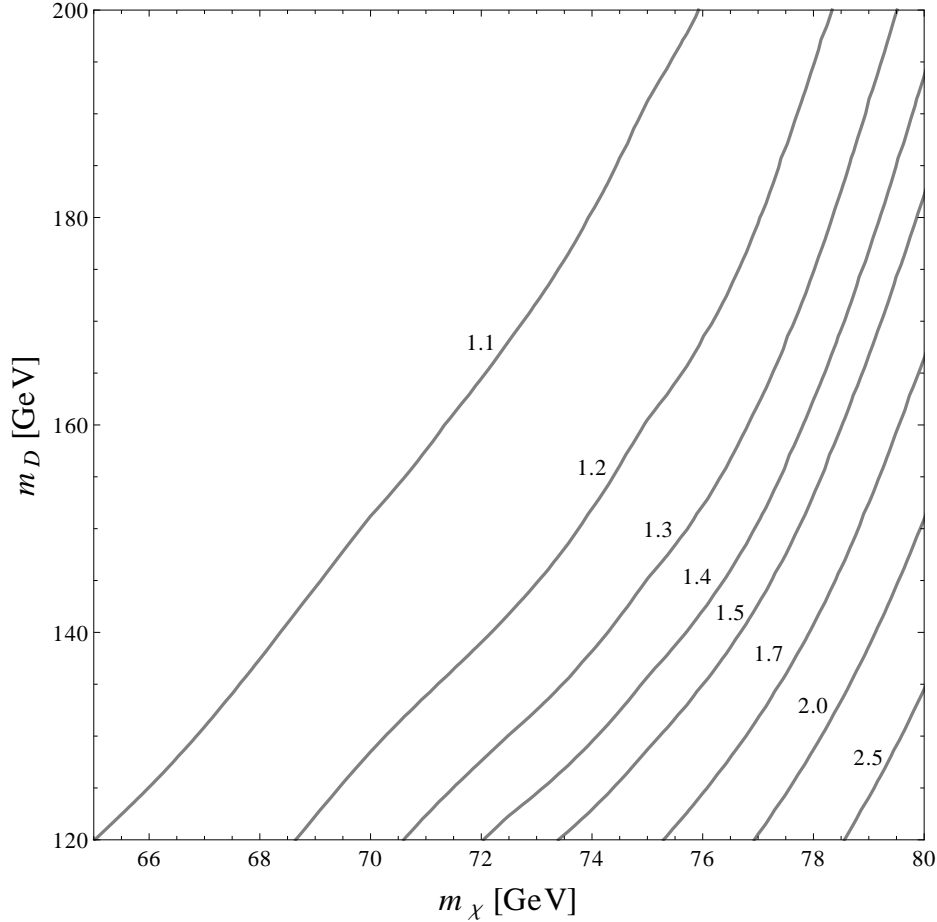}
\end{center}
\caption{\label{fig:WWcontour}Contour plot of $R_{32} \equiv \frac{\Phi^{\text {tot, incl. 3-body}}_\mu}{\Phi^{\text {tot, 2-body}}_\mu} (E_\mu \ge 10 \text{ GeV})$ as a function of $m_\chi = m_N$ and $m_D$ subject to the requirements that $\lambda^\prime = \lambda^\prime_{\text{critical}}$ (such that the dark matter-Higgs boson coupling cancels completely) and that $\lambda$ is fixed by requiring $\Omega h^2 = 0.112$.}
\end{figure}

To give a broader sense of how the importance of annihilation to $WW^\ast$ can vary, we calculate the ratio
\begin{equation}
R_{32} \equiv \frac{\Phi_\mu^{\text{ave, incl. 3-body}}}{\Phi_\mu^{\text{ave, 2-body only}}}(E_\mu \ge E_\mu^{\text{thresh}})
\end{equation}
at DeepCore/IceCube for a variety of points.  Again, this represents the factor by which signals (and hence constraints) are enhanced by including annihilation to $WW^\ast$.  A contour plot showing the $R_{32}$ as a function of $m_\chi = m_N$ and $m_D$ is shown in figure~\ref{fig:WWcontour} for $E_\mu^{\text{thresh}} = 10 \text{ GeV}$, subject to the same requirements as the benchmark point that $\lambda^\prime = \lambda^\prime_{\text{critical}}$ and $\lambda$ is fixed by requiring $\Omega h^2 = 0.112$.  The majority of the variation in the plot arises from changes in the branching ratio to $WW^\ast$.  In general, points of this type ($m_\chi \approx m_W$, suppressed $\sigma_{\text{SI}}$ and $\Omega h^2$ set by annihilation via an $s$-channel $Z$ boson) exhibit the largest values of $\sigma_{\text{SD}}^p$, comparable to current spin-dependent constraints.  \emph{Thus, including the effect of $WW^\ast$ will push neutrino signals above exclusion limits for certain points.}

Given the impact of annihilation to $Wf \bar{f}'$ on neutrino signals, it is reasonable to wonder whether it may also have a significant effect on the relic density.  This would not have been taken into account by our calculations of $\Omega h^2$, which considered only 2-body final states.  The potential importance of 3-body final states on relic density calculations has been highlighted previously in \cite{Yaguna:2010hn}.  However, for the points considered here, the fact that the cross section for $s$-wave annihilation to $WW^\ast$ is comparable to the mass-suppressed cross section for $s$-wave annihilation to fermions suggests that in the early universe $p$-wave annihilation to fermions will still dominate.  This intuition is confirmed by numerical tests.  Comparisons of branching ratios to $W^+ W^-$ in the early universe for $m_\chi = 80.5 \text{ GeV}$ given by \texttt{micrOmegas} indicate the relic density calculation is wrong by at most $\mathcal{O}(10\%)$ for points shown in the contour plot, and that for most points it is much more accurate, to $\mathcal{O}(1\%)$.\footnote{For $m_\chi = 80.5 \text{ GeV} \gsim m_W$, both $W$'s can be on-shell, so the branching ratio to $W^+ W^-$ calculated by considering only 2-body final states will be accurate.  Thus, the difference the branching ratio to $W^+ W^-$ for $m_\chi \gsim m_W$ and $m_\chi \lsim m_W$ gives the approximate error introduced by neglecting annihilation to $WW^\ast$.}
So, the effect of the $W f \bar{f}^\prime$ final state on neutrino signals must be considered even in cases where its effect on the relic density is negligible.

Consistent with the findings in \cite{Wikstrom:2009kw}, we find that the strongest constraints on SD scattering of singlet-doublet dark matter with $60 \text{ GeV} \lsim m_\chi \leq m_W$ arise from indirect detection experiments.  For instance, both bounds given in Eq.~(\ref{eq:revlim}) for $m_\chi = 75 \text{ GeV}$ are more stringent than the direct detection limits from SIMPLE of $\sigma_{\text{SD}}^p < 5.3 \times 10^{-3} \text{ pb}$ \cite{Felizardo:2011uw} and than the preliminary limits from COUPP of $\sigma_{\text{SD}}^p < (4-7) \times 10^{-3} \text{ pb}$ \cite{COUPPproj}.  This motivates a re-evaluation of the constraints on $\sigma_{\text{SD}}^{p}$ in the singlet-doublet model -- the previous analysis performed in \cite{Cohen:2011ec} used direct detection limits from \cite{Felizardo:2011uw} for $m_\chi \leq m_W$.  We perform such a re-evaluation in section~\ref{sec:revisedsdm}, fully taking into account subdominant annihilation channels.


\section{$tt^\ast$}\label{sec:ttast}

In this section, we present general results for dark matter annihilating to $tt^\ast$.  We do not include a discussion of the importance of annihilation to a $tbW$ final state in the singlet-doublet model (as we did for $W f \bar{f}^\prime$) simply because the effects of annihilation to $tt^\ast$ are small in the majority of the singlet-doublet model parameter space.  We comment briefly on this in section~\ref{sec:revisedsdm}.  That said, it is not difficult to construct a model in which annihilation $tt^\ast$ could have a significant effect on neutrino signals.  The heaviness of the top quark obviates the static-limit $(m_f/m_Z)^2$ suppression.  As a result, near threshold, the cross section for annihilation to a $tbW$ final state can readily compete with $b \bar{b}$, $c \bar{c}$ and $\tau^+ \tau^-$.  Furthermore, the neutrino and muon spectra from top quarks and $W$ bosons are significantly harder than those from bottom quarks.  Consequently, as in the case of annihilation to $WW^\ast$, if the dark matter annihilates predominantly to $b \bar{b}$, the combination of these effects might lead to significant enhancement to neutrino signals from annihilation to $tt^\ast$.  For instance, in models with a new $U(1)$ $Z^\prime$, the freedom to control the couplings of quarks to the $Z^\prime$ allows one to essentially make a $tbW$ final state arbitrarily important.  Alternatively, an MSSM model with light stops could allow the $tbW$ final state to dominate, although in such models one must also ensure that the capture rate in the Sun is large enough to give a measurable signal.  Rather than contrive a model to emphasize the potential importance of a $tbW$ final state, we instead present general spectra for $tt^\ast$.  These results can be applied for your favorite model, requiring only a calculation of the relevant branching ratios.

\begin{figure}
\begin{center}
\includegraphics[width=0.75\textwidth]{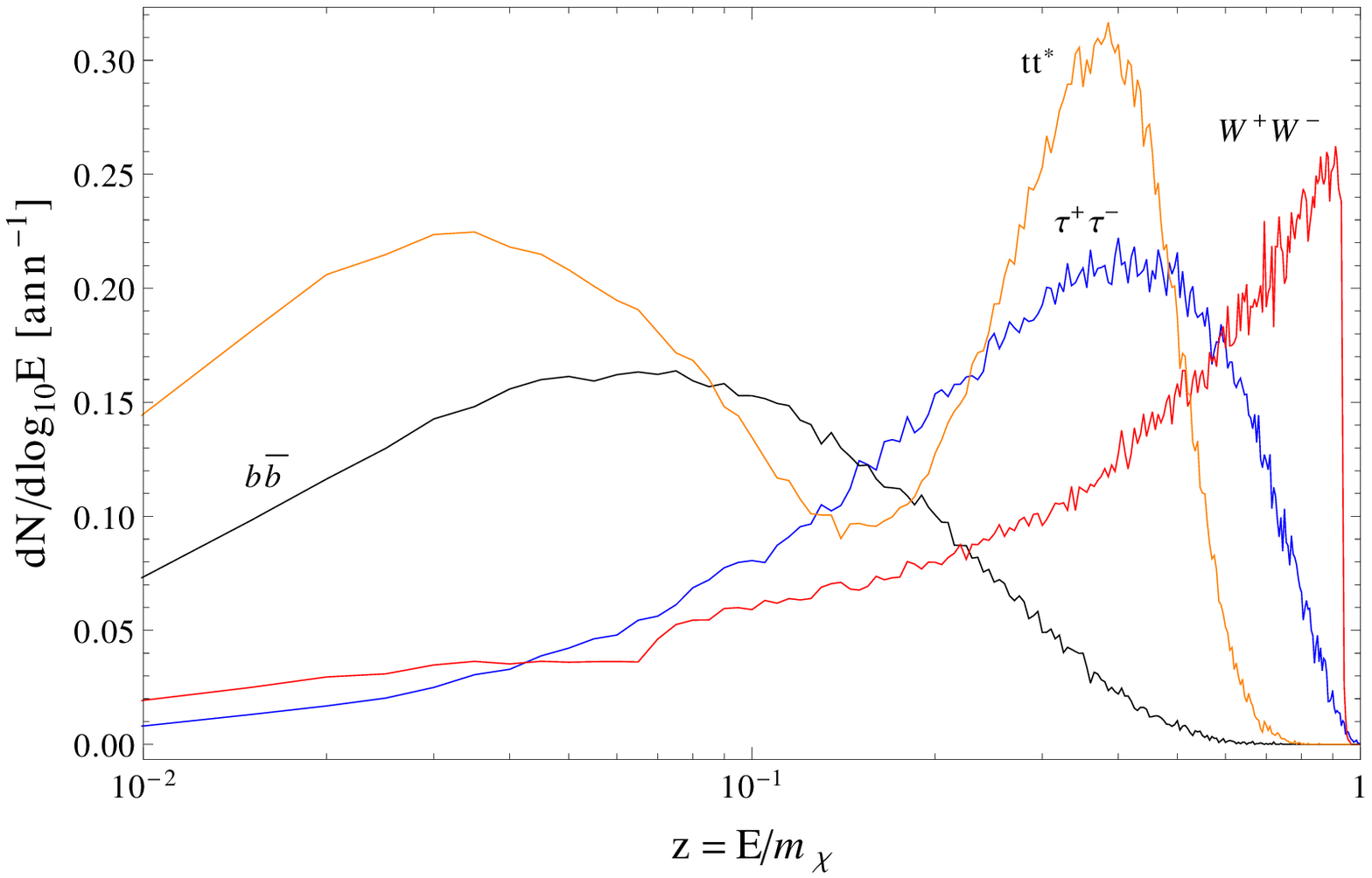}
\includegraphics[width=0.75\textwidth]{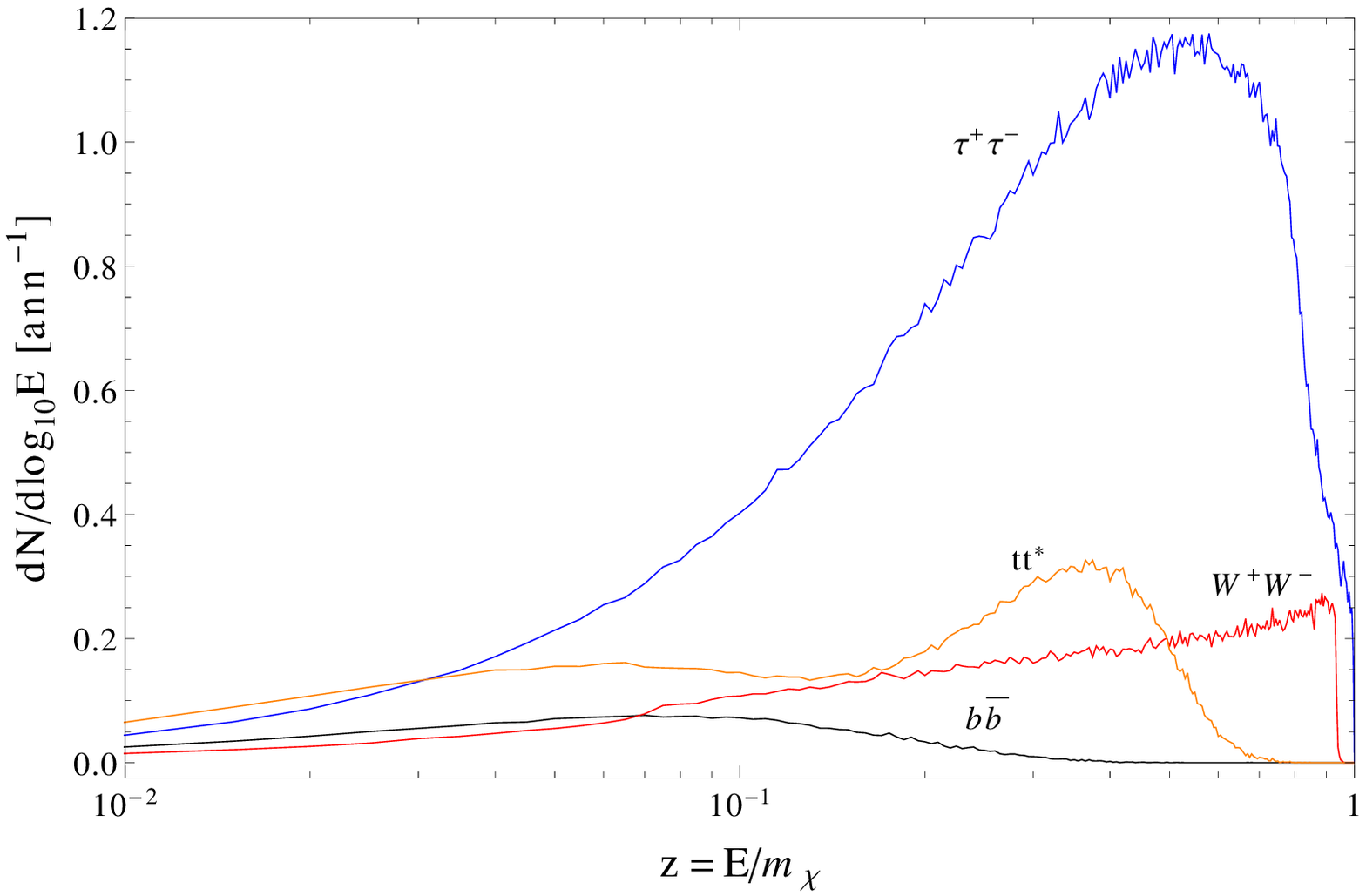}
\end{center}
\caption{\label{fig:ttastinjection}Neutrino injection spectra from dark matter annihilation to $b \bar{b}$ (black), $\tau^+ \tau^-$ (blue), $W^+ W^-$ (red) and $t t^\ast$ (orange) for dark matter mass $m_\chi = 160 \text{ GeV}$.  Shown are the spectra for $\nu_e$ (top) and $\nu_\tau$ (bottom) -- note that $\nu_e = \bar{\nu}_e = \nu_\mu = \bar{\nu}_\mu$ and $\nu_\tau = \bar{\nu}_\tau$.  Annihilation to $t t^\ast$ is assumed to occur via an $s$-channel vector boson.}
\end{figure}

\begin{figure}
\begin{center}
\includegraphics[width=0.75\textwidth]{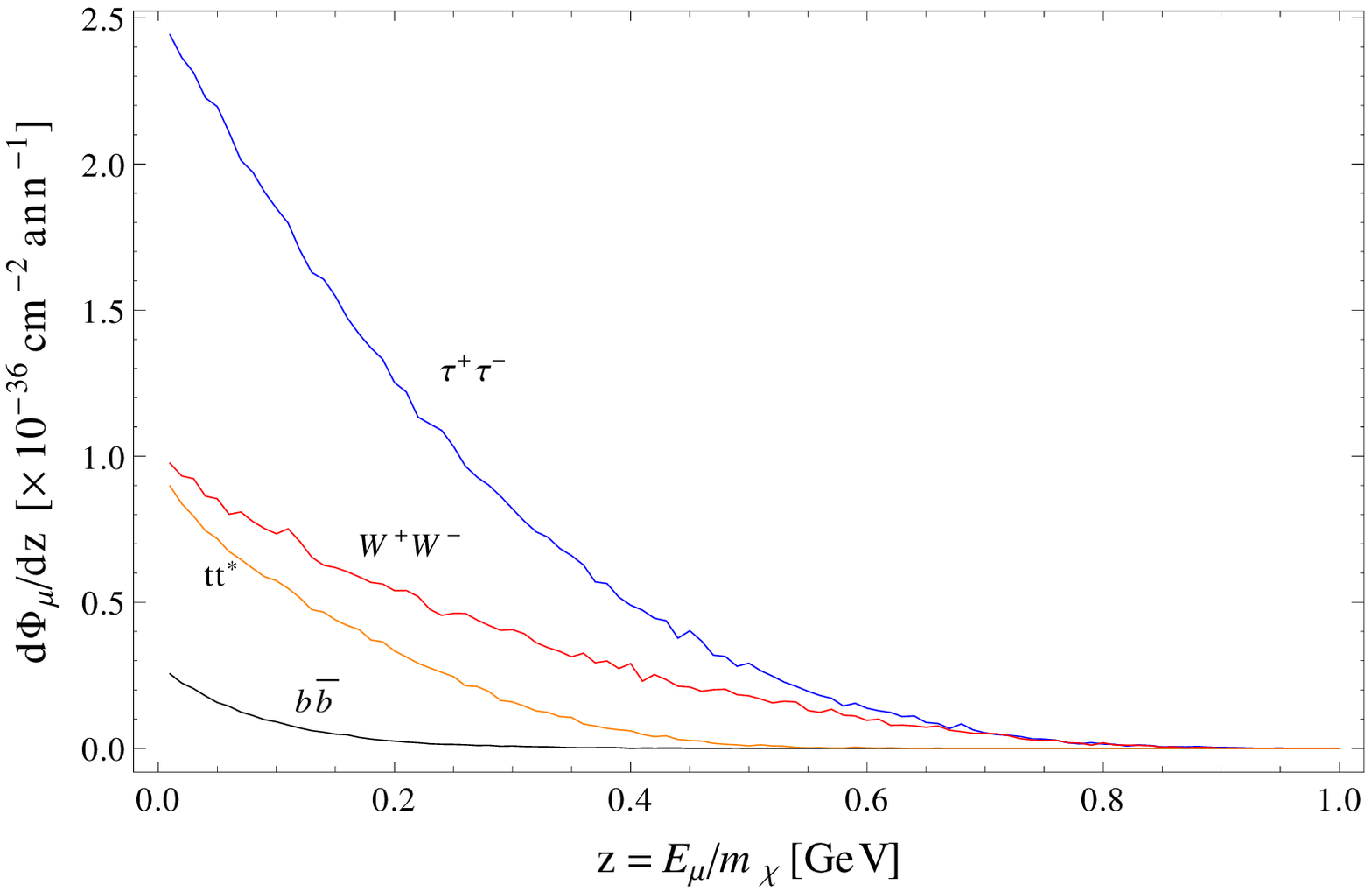}
\end{center}
\caption{\label{fig:ttmuons}Spectra of muons per annihilation at DeepCore/IceCube from annihilation to dark matter annihilation to $b \bar{b}$ (black), $\tau^+ \tau^-$ (blue), $W^+ W^-$ (red) and $t t^\ast$ (orange) for dark matter mass $m_\chi = 160 \text{ GeV}$.  Annihilation to $t t^\ast$ is again assumed to occur via an $s$-channel vector boson.  Type of flux is the same as that of figure~\ref{fig:WWmuons}.}
\end{figure}

The neutrino injection spectra from Majorana dark matter with $m_\chi = 160 \text{ GeV}$ annihilating to a $tbW$ final state via an $s$-channel massive, neutral, vector boson are shown in figure~\ref{fig:ttastinjection}.  Also shown are spectra from $b \bar{b}$, $\tau^+ \tau^-$ and $W^+ W^-$ 2-body annihilation final states from WimpSim.  Spectra from annihilation to $ZZ$ are comparable to those from $W^+ W^-$, so are omitted for clarity.  The corresponding muon spectra at DeepCore/IceCube are shown in figure~\ref{fig:ttmuons}.  The integrated number of muons above threshold per annihilation for two different thresholds, $E_\mu^{\text{thresh}} = 10 \text{ GeV}$ and $E_\mu^{\text{thresh}} = 35 \text{ GeV}$, are given in table~\ref{tab:tbwflux}, and ratios of muon flux from $tbW$ to fluxes from the various 2-body final states are given in table~\ref{tab:nttast}.  Ratios of muon fluxes (and neutrino and muon spectra) are largely constant over the range $150 \text{ GeV} \leq m_\chi \leq m_t$, although variation in neutrino energy with dark matter mass can lead to changes of $\mathcal{O}(10-20\%)$ in integrated fluxes at either end of the range.

\begin{table}
\setstretch{1.5}
\begin{center}
\begin{tabular}{| @{\hspace{0.2cm}} c @{\hspace{0.2cm}} | @{\hspace{0.2cm}} c @{\hspace{0.2cm}} | @{\hspace{0.2cm}} c @{\hspace{0.2cm}} |}
\hline
Final State & $\Phi_\mu^{\text{final state}} (E_\mu \ge 10 \text{ GeV})$ & $\Phi_\mu^{\text{final state}} (E_\mu \ge 35 \text{ GeV})$ \\ \hline
$b \bar{b}$ & $1.0 \times 10^{-38}$ & $1.5 \times 10^{-39}$ \\
$\tau^+ \tau^-$ & $4.6 \times 10^{-37}$ & $2.2 \times 10^{-37}$ \\
$W^+ W^-$ & $2.2 \times 10^{-37}$ & $1.2 \times 10^{-37}$ \\
$t b W$ & $1.0 \times 10^{-37}$ & $3.3 \times 10^{-38}$ \\
\hline
\end{tabular}
\end{center}
\setstretch{1}
\caption{\label{tab:tbwflux}Flux of muons $[\text{cm}^{-2} \text{ ann}^{-1}]$ with energy $E_\mu \ge E_\mu^{\text{thresh}}$ at DeepCore/IceCube from annihilation of dark matter with $m_\chi = 160 \text{ GeV}$ to various final states, including to $tbW$ via an $s$-channel vector boson.}
\end{table}

\begin{table}
\setstretch{1.5}
\begin{center}
\begin{tabular}{| @{\hspace{0.2cm}} l @{\hspace{0.2cm}} | @{\hspace{0.2cm}} c @{\hspace{0.2cm}} | @{\hspace{0.2cm}} c @{\hspace{0.2cm}} |}
\hline
 & $E_\mu^{\text{thresh}} = 10 \text{ GeV}$ & $E_\mu^{\text{thresh}} = 35 \text{ GeV}$ \\ \hline
$\Phi^{tbW}_\mu/\Phi^{\tau^+ \tau^-}_\mu$ & 0.22 & 0.15 \\
$\Phi^{tbW}_\mu/\Phi^{W^+ W^-}_\mu$ & 0.48 & 0.28 \\
$\Phi^{tbW}_\mu/\Phi^{b \bar{b}}_\mu$ & 9.6 & 22 \\ \hline
\end{tabular}
\end{center}
\setstretch{1}
\caption{\label{tab:nttast}Ratios of total number of muons per annihilation above two different threshold energies for annihilation of dark matter with mass $m_\chi = 160 \text{ GeV}$ to various final states, including to $tbW$ via an $s$-channel vector boson.}
\end{table}

Once again, the hardest neutrino and muon spectra are produced by annihilation to $\tau^+ \tau^-$.  Muons from $tbW$ are also slightly softer than those from $W^+ W^-$, but are significantly harder than those from $b \bar{b}$.
Thus, annihilation to $tbW$ may enhance indirect detection signals if the dominant 2-body annihilation mode is to $b \bar{b}$ or other light fermions (which tend to have even softer spectra than $b \bar{b}$, again with the exception of $\tau^+ \tau^-$).   
It may also be the case that the possibility of annihilation to $tbW$ degrades indirect detection signals for models in which the dominant contributions to the muon flux arise from annihilation to $\tau^+ \tau^-$ or $W^+ W^-$ (as annihilation to $tbW$ may decrease the branching ratio to these final states).

\section{Revised Indirect Detection Constraints and Discovery Prospects for the Singlet-Doublet Model}\label{sec:revisedsdm}

In section~\ref{sec:wwastsdm}, we found that the indirect detection limits from Super-K on singlet-doublet dark matter with $m_\chi \leq m_W$ were significantly more stringent than those quoted in \cite{Tanaka:2011uf}, and stronger than the direct detection limits from COUPP and SIMPLE.  Indirect detection limits are also stronger than those from collider experiments.\footnote{ATLAS mono-jet and missing energy searches \cite{ATLASmonojet} can place robust bounds on dark matter-quark interactions (see, for instance, \cite{Goodman:2010ku, Fox:2011pm, Rajaraman:2011wf}) but, because the mediator of singlet-doublet dark matter-quark interactions is the (relatively) light $Z$ boson, collider limits are substantially weakened in this model \cite{Goodman:2011jq}.} This merits a revision of the indirect detection limits on the singlet-doublet model presented in \cite{Cohen:2011ec}.

\subsection{$m_\chi < m_W$}

Figure~\ref{fig:WWcontour} demonstrates that, for $m_\chi \lsim m_W$, annihilation to $WW^\ast$ can enhance neutrino signals for specific points in the singlet-doublet model. 
However, even if the looser bound $\sigma_{\text{SD}}^{p,\text{2-body only}} < 2.4 \times 10^{-3} \text{ pb}$ for singlet-doublet dark matter with $m_\chi \approx 75 \text{ GeV}$ is applied, we find that this minimal model is more tightly constrained than suggested in \cite{Cohen:2011ec}, which used the SIMPLE bound of $\sigma_{\text{SD}}^p \lsim 5 \times 10^{-3} \text{ pb}$.  In fact, this bound has significant implications for Majorana dark matter in general.  For Majorana dark matter with $m_\chi < m_W$ and thermal relic abundance set by annihilation via an $s$-channel $Z$ boson, $\sigma_{\text{SD}}^p$ is fixed for a particular value of $m_\chi$.   The reason is simple: the dark matter-$Z$ boson coupling is set by requiring the correct relic density -- see, for example, figure 5 of \cite{Cohen:2011ec} and discussion thereof.  
Dark matter annihilation in the Sun also occurs via $s$-channel $Z$ boson exchange.  So, for masses somewhat below $m_W$, the ``2-body only" branching ratios given in table~\ref{tab:brwwast}, and thus the bound, apply.  
We find that
\emph{this bound excludes Majorana dark matter with $70 \text{ GeV} \lsim m_\chi \lsim m_W$ whose annihilations are controlled by an $s$-channel Z}.

For $m_\chi \lsim 70 \text{ GeV}$, the proximity of $m_\chi$ to $\frac{m_Z}{2}$ causes dark matter annihilation via an $s$-channel $Z$ boson in the early universe to be enhanced due to the smaller propagator.  As a result, the correct relic density can be achieved with a smaller dark matter-$Z$ boson coupling, leading to suppressed values of $\sigma_{\text{SD}}^p$ that still evade this bound.  However, in the singlet-doublet model specifically, the enhancement to neutrino signals from annihilation to $WW^\ast$ can still lead to the exclusion of some points with $m_\chi \lsim 70 \text{ GeV}$ that exhibit a non-negligible branching ratio to $WW^\ast$.  This effect is most relevant for small $m_{D}$.

\subsection{$m_W < m_\chi < m_t$}

For $m_W < m_\chi < m_t$, the main contribution to neutrino signals in the singlet-doublet model generally comes from annihilation to on-shell electroweak boson pairs.  In the static limit, the dark matter can annihilate to $W^+ W^-$ via $t$- and $u$-channel exchange of the charged $SU(2)$-partner of the dark matter or, for $m_\chi > m_Z$, to $Z Z$ via $t$- and $u$-channel exchange of the dark matter itself and the heavier neutral states, $\nu_2$ and $\nu_3$.  If $2 m_\chi > m_Z + m_h$, annihilation to a $Z h$ final state, both via $t$- and $u$-channel exchange of the neutral states and via an $s$-channel $Z$ boson, will also occur.  Assuming annihilation exclusively to $W^+ W^-$ would correspond to the ``hard'' limit quoted in \cite{Tanaka:2011uf}, which is
\begin{equation}
\label{eq:strongbound}
\sigma_{\text{SD}}^{p} < 3.0 \times 10^{-4} \text{ pb}  \rm{\, \, (hard, \, \,} W^+W^- \rm{\, \, or\, \,}  ZZ\rm{)}
\end{equation}
for $m_\chi = 100 \text{ GeV}$.  This is a factor of $\sim 10$ stronger than the corresponding previous bound quoted in \cite{Desai:2004pq} (used in \cite{Cohen:2011ec}).  For larger values of $m_\chi$, the decrease in the local number density $n_\chi = \rho/m_\chi$ weakens the bound, whereas the increase in $\Phi_\mu^{W^+ W^-} (E_\mu \ge 2 \text{ GeV})$ strengthens it.  These effects are comparable, such that the hard limit is $\sigma_{\text{SD}}^{p} < (3-4) \times 10^{-4} \text{ pb}$ over the entire range $m_W \lsim m_\chi \lsim m_t$.

How does this compare to the typical values of $\sigma_{\text{SD}}^p$ in the singlet-doublet model?  Singlet-doublet dark matter with mass $m_W < m_\chi < m_t$, which does not undergo co-annihilation in the early universe (and evades bounds on $\sigma_{\text{SI}}$ from XENON100 \cite{Aprile:2008rc}), achieves the correct relic density predominantly by annihilation via an $s$-channel $Z$ to $W^+ W^-$, light fermions and (if possible) $Z h$, with additional contributions from annihilation to electroweak boson pairs via $t$- and $u$-channel fermion exchange.  As such, achieving the correct relic density requires a small range of  dark matter-$Z$ boson coupling.  This fixes $\sigma_{\text{SD}}^p \gsim 2 \times 10^{-3} \text{ pb}$ for singlet-doublet dark matter of this type (the upper horizontal band of figure 4 in \cite{Cohen:2011ec}).  The hard limit of Eq.~(\ref{eq:strongbound}) is a factor of $\sim 7$ below this, \emph{which would exclude all such points}.  For a light Higgs boson ($m_h \lsim 140 \text{ GeV}$ -- favored by recent ATLAS and CMS results \cite{atlashiggs, cmshiggs}) and $m_W < m_\chi < m_t$, the situation is becoming squeezed for singlet-doublet dark matter: to avoid the combination of SI and SD experimental constraints, dark sector masses must be tuned to permit early universe co-annihilation.  Only in this case can the correct relic density be achieved without the corresponding generation of large dark matter-nucleon scattering cross sections.


While these conclusions are based on the hard limit given in Eq.~(\ref{eq:strongbound}), whereas exact bounds depend on the specific annihilation branching ratios, we nonetheless find them to be robust.
The $ZZ$ and $W^+ W^-$ final states yield sufficiently comparable spectra that annihilation to $ZZ$ instead of $W^+ W^-$ would not significantly alter the bound.  Annihilation to $Zh$ would weaken the bound by at worst a factor of $2$ (assuming the Higgs boson is sufficiently light that it decays overwhelmingly to $b \bar{b}$, which would contribute negligibly to indirect detection signals relative to the single $Z$).  The bound would weaken more drastically if annihilation to light fermions could be made to dominate.  However, this is not easy to do.  While annihilation to $W^+ W^-$ is suppressed in the static limit for large $m_D$, the same cannot be said of annihilation to $Z Z$, for which the mass of one of the exchanged particles is fixed to be $m_\chi$.  Thus, if the dark matter-$Z$ boson coupling is large enough to generate a sizable $\sigma_{\text{SD}}^p$, there will generically be a sizable cross section for dark matter annihilation to $ZZ$.  Furthermore, potentially competing cross sections for annihilation to light fermions are suppressed by $(m_f/m_Z)^2$.  Consequently, the branching ratio for annihilation to boson pairs invariably dominates, and we find that overall neutrino signals are at most degraded by a factor of $\sim 2$.  As the bounds are significantly lower (a factor of $\sim 7$) than the general $\sigma_{\text{SD}}^p$ of interest in the singlet-doublet model, such a degradation would not affect the conclusion that the majority of points with large $\sigma_{\text{SD}}^p$ and $m_W < m_\chi < m_t$ are excluded.  In fact, the bound is sufficiently strong that this holds even if the bound is weakened by taking more pessimistic (but reasonable) choices of astrophysical parameters.

\subsection{$m_\chi > m_t$}

In this region of parameter space, unsuppressed $s$-wave annihilation to $t \bar{t}$ via an $s$-channel $Z$ boson allows the correct relic density to be achieved with a significantly lower value of $\sigma_{\text{SD}}^p$ -- singlet-doublet dark matter with $m_\chi > m_t$ that evades bounds on $\sigma_{\text{SI}}$ from XENON100 typically exhibits $\sigma_{\text{SD}}^p \gsim 7 \times 10^{-5} \text{ pb}$ (the lower horizontal band of figure 4 in \cite{Cohen:2011ec}).  As a result, current bounds do not yet constrain $\sigma_{\text{SD}}^p$ for singlet-doublet dark matter with $m_\chi > m_t$.  The situation is more optimistic for DeepCore; the projected DeepCore limits are $\sigma_{\text{SD}}^p \lsim (2-8) \times 10^{-5} \text{ pb}$ for annihilation of dark matter with $m_t < m_\chi \lsim 800 \text{ GeV}$ to $W^+ W^-$ \cite{DeYoung:2011ke}.  The hardest neutrinos from the $t \bar{t}$ final state arise from the $W$'s produced in $t \rightarrow bW$ decay, which will be softer than the $W$'s produced in direct $\chi \chi \rightarrow W^+ W^-$ annihilations.  Consequently, the $t \bar{t}$ spectra are softer than those from the $W^+ W^-$ final state and the comparable $Z Z$ final state (as one would have expected from the results of section~\ref{sec:ttast}).  Thus, for singlet-doublet dark matter with $m_\chi > m_t$ and large values of $\sigma_{\text{SD}}^p$ (which habitually annihilates to $t \bar{t}$ via an $s$-channel $Z$ boson in the static limit), table~\ref{tab:nttast} indicates that the projected hard DeepCore limits will be degraded by a factor of $\sim 2 \; (4)$ for a threshold of $E_\mu^{\text{thresh}} = 10 \; (35) \text{ GeV}$.  However, supposing limits comparable to those projected are achieved, DeepCore will still be sufficiently sensitive to probe much of the remaining singlet-doublet parameter space with $m_\chi > m_t$, consistent with the claims of \cite{Cohen:2011ec}.

As limits derived from assuming annihilation to $t \bar{t}$ are weaker than those for annihilation to $W^+ W^-$, one might worry that annihilation to a $tbW$ final state could degrade the indirect detection limits just discussed for $m_\chi \lsim m_t$.  However, for Super-K the degradation would be at worst a factor of $\sim 2$ (if the dark matter annihilated exclusively to $t t^\ast$) due to Super-K's low $E_\mu^{\text{thresh}}$.  As discussed in the last subsection, this degradation would not change the conclusion that the majority of points with large $\sigma_{\text{SD}}^p$ and $m_W < m_\chi < m_t$ are excluded.
Assuming that that the hard limits are degraded by a factor of $\sim 2$, we can ask at what point the contribution from annihilation to $tt^\ast$ in the early universe is sufficiently large to permit the right relic density to be achieved with a value of $\sigma_{\text{SD}}^p$ that is small enough to avoid the (degraded) bound.  In other words, there may be exceptional points with $m_\chi \lsim m_t$ for which SI constraints are evaded and a thermal relic density is achieved, which are not excluded by Super-K due to a combination of two factors:
\begin{enumerate}
\item The correct thermal relic density is achieved with a smaller dark matter-$Z$ boson coupling (and hence a smaller $\sigma_{\text{SD}}^p$) 
due to the significant contribution from annihilation to $tt^\ast$ in the early universe.
\item Hard indirect detection limits are degraded by a factor of $\sim 2$ due to the dark matter annihilating predominantly to $tt^\ast$ in the Sun.  For $m_\chi \lsim m_t$, the hard limit from Super-K is $\sigma_{\text{SD}}^p < 4.0 \times 10^{-4} \text{ pb}$ \cite{Tanaka:2011uf}, leading to a degraded limit of $\sigma_{\text{SD}}^p \lsim 8.0 \times 10^{-4} \text{ pb}$. 
\end{enumerate}
 Numerically integrating expressions for $\langle{\sigma v\rangle}$ and $\Omega h^2$ (from \cite{Gondolo:1990dk}), we find that these conditions are only fulfilled for  $m_\chi$ approximately $2 - 3 \text{ GeV}$ less than $m_t$.  This indicates that Super-K, when taken in concert with XENON100, indeed excludes much of the singlet-doublet parameter space from $m_{\chi} \approx 70$ GeV all the way up to $m_\chi \approx m_t$.

\section{Conclusion}\label{sec:conclusion}

In this work, we have explored the potential importance of dark matter annihilation to 3-body final states near threshold, and have highlighted situations in which certain such final states can significantly impact the reach of indirect detection experiments.  In particular, we have shown that for the minimal singlet-doublet model of dark matter, consideration of the $W f \bar{f}^\prime$ final state can improve indirect detection bounds by up to a factor of $\sim (1.5 - 2)$ for $m_\chi \lsim m_W$.  
Meanwhile, annihilation to $t b W$ will frequently degrade bounds by competing with annihilation to electroweak boson pairs, but this turns out to be largely unimportant in the singlet-doublet model except for $m_\chi$ very close to $m_t$.  

We have also demonstrated that indirect detection searches (performed by Super-K and IceCube) still provide the most extensive probe of spin-dependent dark matter-nucleon scattering over much of the singlet-doublet model parameter space, though direct detection is becoming competitive.  
In fact, for $m_\chi \lsim m_W$, bounds from direct and indirect detection experiments are extremely similar for dark matter annihilating exclusively to $b \bar{b}$ in the Sun.  Indirect detection bounds are more stringent for singlet-doublet dark matter with $m_\chi \lsim m_W$ because of the sizable contribution to neutrino signals from the subdominant but hard $\tau^+ \tau^-$ channel, which doubles the strength of indirect detection bounds in spite of its small ($\sim 5\%$) branching ratio.
Data from DeepCore will prove conclusive for the singlet-doublet model with a thermal history, except perhaps in the exceptional cases --  the dark matter sits on resonance, or co-annihilation is crucial in setting the relic density.

These findings reinforce the sensitivity of neutrino signals to dark matter annihilation branching ratios.
As such, signals may encode information about other Beyond the Standard Model particles -- for instance, the effect of the $W f \bar{f}^\prime$ final state in the singlet-doublet model is sensitive to the mass of the charged fermion.  Thus, an understanding of 3-body final states may be important to maximizing the utility of indirect detection experiments. 

\vspace{-0.2cm}
\subsection*{Acknowledgments}
\vspace{-0.3cm}
We thank Timothy Cohen and David Tucker-Smith
for useful conversations and comments on the draft.
J.K. is particularly thankful to Ran Lu for introduction to and assistance with several pieces of software used in this analysis.
The work of A.P. was supported in part by NSF Career Grant NSF-PHY-0743315 and by DOE Grant \#DE-FG02-95ER40899.

%

\bibliography{./NeutrinoBib}

\begin{thebibliography}{50}%
\makeatletter
\providecommand \@ifxundefined [1]{%
 \@ifx{#1\undefined}
}%
\providecommand \@ifnum [1]{%
 \ifnum #1\expandafter \@firstoftwo
 \else \expandafter \@secondoftwo
 \fi
}%
\providecommand \@ifx [1]{%
 \ifx #1\expandafter \@firstoftwo
 \else \expandafter \@secondoftwo
 \fi
}%
\providecommand \natexlab [1]{#1}%
\providecommand \enquote  [1]{``#1''}%
\providecommand \bibnamefont  [1]{#1}%
\providecommand \bibfnamefont [1]{#1}%
\providecommand \citenamefont [1]{#1}%
\providecommand \href@noop [0]{\@secondoftwo}%
\providecommand \href [0]{\begingroup \@sanitize@url \@href}%
\providecommand \@href[1]{\@@startlink{#1}\@@href}%
\providecommand \@@href[1]{\endgroup#1\@@endlink}%
\providecommand \@sanitize@url [0]{\catcode `\\12\catcode `\$12\catcode
  `\&12\catcode `\#12\catcode `\^12\catcode `\_12\catcode `\%12\relax}%
\providecommand \@@startlink[1]{}%
\providecommand \@@endlink[0]{}%
\providecommand \url  [0]{\begingroup\@sanitize@url \@url }%
\providecommand \@url [1]{\endgroup\@href {#1}{\urlprefix }}%
\providecommand \urlprefix  [0]{URL }%
\providecommand \Eprint [0]{\href }%
\providecommand \doibase [0]{http://dx.doi.org/}%
\providecommand \selectlanguage [0]{\@gobble}%
\providecommand \bibinfo  [0]{\@secondoftwo}%
\providecommand \bibfield  [0]{\@secondoftwo}%
\providecommand \translation [1]{[#1]}%
\providecommand \BibitemOpen [0]{}%
\providecommand \bibitemStop [0]{}%
\providecommand \bibitemNoStop [0]{.\EOS\space}%
\providecommand \EOS [0]{\spacefactor3000\relax}%
\providecommand \BibitemShut  [1]{\csname bibitem#1\endcsname}%
\let\auto@bib@innerbib\@empty
\bibitem [{\citenamefont {Press}\ and\ \citenamefont
  {Spergel}(1985)}]{Press:1985ug}%
  \BibitemOpen
  \bibfield  {author} {\bibinfo {author} {\bibfnamefont {W.~H.}\ \bibnamefont
  {Press}}\ and\ \bibinfo {author} {\bibfnamefont {D.~N.}\ \bibnamefont
  {Spergel}},\ }\href {\doibase 10.1086/163485} {\bibfield  {journal} {\bibinfo
   {journal} {Astrophys. J.}\ }\textbf {\bibinfo {volume} {296}},\ \bibinfo
  {pages} {679} (\bibinfo {year} {1985})}\BibitemShut {NoStop}%
\bibitem [{\citenamefont {Silk}\ \emph {et~al.}(1985)\citenamefont {Silk},
  \citenamefont {Olive},\ and\ \citenamefont {Srednicki}}]{Silk:1985ax}%
  \BibitemOpen
  \bibfield  {author} {\bibinfo {author} {\bibfnamefont {J.}~\bibnamefont
  {Silk}}, \bibinfo {author} {\bibfnamefont {K.~A.}\ \bibnamefont {Olive}}, \
  and\ \bibinfo {author} {\bibfnamefont {M.}~\bibnamefont {Srednicki}},\ }\href
  {\doibase 10.1103/PhysRevLett.55.257} {\bibfield  {journal} {\bibinfo
  {journal} {Phys. Rev. Lett.}\ }\textbf {\bibinfo {volume} {55}},\ \bibinfo
  {pages} {257} (\bibinfo {year} {1985})}\BibitemShut {NoStop}%
\bibitem [{\citenamefont {Srednicki}\ \emph {et~al.}(1987)\citenamefont
  {Srednicki}, \citenamefont {Olive},\ and\ \citenamefont
  {Silk}}]{Srednicki:1986vj}%
  \BibitemOpen
  \bibfield  {author} {\bibinfo {author} {\bibfnamefont {M.}~\bibnamefont
  {Srednicki}}, \bibinfo {author} {\bibfnamefont {K.~A.}\ \bibnamefont
  {Olive}}, \ and\ \bibinfo {author} {\bibfnamefont {J.}~\bibnamefont {Silk}},\
  }\href {\doibase 10.1016/0550-3213(87)90020-4} {\bibfield  {journal}
  {\bibinfo  {journal} {Nucl. Phys.}\ }\textbf {\bibinfo {volume} {B279}},\
  \bibinfo {pages} {804} (\bibinfo {year} {1987})}\BibitemShut {NoStop}%
\bibitem [{\citenamefont {Jungman}\ \emph {et~al.}(1996)\citenamefont
  {Jungman}, \citenamefont {Kamionkowski},\ and\ \citenamefont
  {Griest}}]{Jungman:1995df}%
  \BibitemOpen
  \bibfield  {author} {\bibinfo {author} {\bibfnamefont {G.}~\bibnamefont
  {Jungman}}, \bibinfo {author} {\bibfnamefont {M.}~\bibnamefont
  {Kamionkowski}}, \ and\ \bibinfo {author} {\bibfnamefont {K.}~\bibnamefont
  {Griest}},\ }\href {\doibase 10.1016/0370-1573(95)00058-5} {\bibfield
  {journal} {\bibinfo  {journal} {Phys.Rept.}\ }\textbf {\bibinfo {volume}
  {267}},\ \bibinfo {pages} {195} (\bibinfo {year} {1996})},\ \Eprint
  {http://arxiv.org/abs/hep-ph/9506380} {arXiv:hep-ph/9506380 [hep-ph]}
  \BibitemShut {NoStop}%
\bibitem [{\citenamefont {Kamionkowski}\ \emph {et~al.}(1995)\citenamefont
  {Kamionkowski}, \citenamefont {Griest}, \citenamefont {Jungman},\ and\
  \citenamefont {Sadoulet}}]{Kamionkowski:1994dp}%
  \BibitemOpen
  \bibfield  {author} {\bibinfo {author} {\bibfnamefont {M.}~\bibnamefont
  {Kamionkowski}}, \bibinfo {author} {\bibfnamefont {K.}~\bibnamefont
  {Griest}}, \bibinfo {author} {\bibfnamefont {G.}~\bibnamefont {Jungman}}, \
  and\ \bibinfo {author} {\bibfnamefont {B.}~\bibnamefont {Sadoulet}},\ }\href
  {\doibase 10.1103/PhysRevLett.74.5174} {\bibfield  {journal} {\bibinfo
  {journal} {Phys.Rev.Lett.}\ }\textbf {\bibinfo {volume} {74}},\ \bibinfo
  {pages} {5174} (\bibinfo {year} {1995})},\ \Eprint
  {http://arxiv.org/abs/hep-ph/9412213} {arXiv:hep-ph/9412213 [hep-ph]}
  \BibitemShut {NoStop}%
\bibitem [{\citenamefont {Wikstrom}\ and\ \citenamefont
  {Edsjo}(2009)}]{Wikstrom:2009kw}%
  \BibitemOpen
  \bibfield  {author} {\bibinfo {author} {\bibfnamefont {G.}~\bibnamefont
  {Wikstrom}}\ and\ \bibinfo {author} {\bibfnamefont {J.}~\bibnamefont
  {Edsjo}},\ }\href {\doibase 10.1088/1475-7516/2009/04/009} {\bibfield
  {journal} {\bibinfo  {journal} {JCAP}\ }\textbf {\bibinfo {volume} {0904}},\
  \bibinfo {pages} {009} (\bibinfo {year} {2009})},\ \Eprint
  {http://arxiv.org/abs/0903.2986} {arXiv:0903.2986 [astro-ph.CO]} \BibitemShut
  {NoStop}%
\bibitem [{\citenamefont {DeYoung}(2011)}]{DeYoung:2011ke}%
  \BibitemOpen
  \bibfield  {author} {\bibinfo {author} {\bibfnamefont {T.}~\bibnamefont
  {DeYoung}} (\bibinfo {collaboration} {for the IceCube Collaboration}),\
  }\href@noop {} {\  (\bibinfo {year} {2011})},\ \Eprint
  {http://arxiv.org/abs/1112.1053} {arXiv:1112.1053 [astro-ph.HE]} \BibitemShut
  {NoStop}%
\bibitem [{\citenamefont {Ha}(2012)}]{Ha:2012ww}%
  \BibitemOpen
  \bibfield  {author} {\bibinfo {author} {\bibfnamefont {C.~H.}\ \bibnamefont
  {Ha}} (\bibinfo {collaboration} {for the IceCube Collaboration}),\
  }\href@noop {} {\  (\bibinfo {year} {2012})},\ \Eprint
  {http://arxiv.org/abs/1201.0801} {arXiv:1201.0801 [hep-ex]} \BibitemShut
  {NoStop}%
\bibitem [{\citenamefont {Cirelli}\ \emph {et~al.}(2005)\citenamefont
  {Cirelli}, \citenamefont {Fornengo}, \citenamefont {Montaruli}, \citenamefont
  {Sokalski}, \citenamefont {Strumia} \emph {et~al.}}]{Cirelli:2005gh}%
  \BibitemOpen
  \bibfield  {author} {\bibinfo {author} {\bibfnamefont {M.}~\bibnamefont
  {Cirelli}}, \bibinfo {author} {\bibfnamefont {N.}~\bibnamefont {Fornengo}},
  \bibinfo {author} {\bibfnamefont {T.}~\bibnamefont {Montaruli}}, \bibinfo
  {author} {\bibfnamefont {I.~A.}\ \bibnamefont {Sokalski}}, \bibinfo {author}
  {\bibfnamefont {A.}~\bibnamefont {Strumia}},  \emph {et~al.},\ }\href
  {\doibase 10.1016/j.nuclphysb.2005.08.017, 10.1016/j.nuclphysb.2007.10.001}
  {\bibfield  {journal} {\bibinfo  {journal} {Nucl.Phys.}\ }\textbf {\bibinfo
  {volume} {B727}},\ \bibinfo {pages} {99} (\bibinfo {year} {2005})},\ \Eprint
  {http://arxiv.org/abs/hep-ph/0506298} {arXiv:hep-ph/0506298 [hep-ph]}
  \BibitemShut {NoStop}%
\bibitem [{\citenamefont {Blennow}\ \emph {et~al.}(2008)\citenamefont
  {Blennow}, \citenamefont {Edsjo},\ and\ \citenamefont
  {Ohlsson}}]{Blennow:2007tw}%
  \BibitemOpen
  \bibfield  {author} {\bibinfo {author} {\bibfnamefont {M.}~\bibnamefont
  {Blennow}}, \bibinfo {author} {\bibfnamefont {J.}~\bibnamefont {Edsjo}}, \
  and\ \bibinfo {author} {\bibfnamefont {T.}~\bibnamefont {Ohlsson}},\ }\href
  {\doibase 10.1088/1475-7516/2008/01/021} {\bibfield  {journal} {\bibinfo
  {journal} {JCAP}\ }\textbf {\bibinfo {volume} {0801}},\ \bibinfo {pages}
  {021} (\bibinfo {year} {2008})},\ \Eprint {http://arxiv.org/abs/0709.3898}
  {arXiv:0709.3898 [hep-ph]} \BibitemShut {NoStop}%
\bibitem [{\citenamefont {Chen}\ and\ \citenamefont
  {Kamionkowski}(1998)}]{Chen:1998pp}%
  \BibitemOpen
  \bibfield  {author} {\bibinfo {author} {\bibfnamefont {X.-l.}\ \bibnamefont
  {Chen}}\ and\ \bibinfo {author} {\bibfnamefont {M.}~\bibnamefont
  {Kamionkowski}},\ }\href@noop {} {\ ,\ \bibinfo {pages} {359} (\bibinfo
  {year} {1998})},\ \Eprint {http://arxiv.org/abs/hep-ph/9901435}
  {arXiv:hep-ph/9901435 [hep-ph]} \BibitemShut {NoStop}%
\bibitem [{\citenamefont {Bergstrom}(1989)}]{Bergstrom:1989jr}%
  \BibitemOpen
  \bibfield  {author} {\bibinfo {author} {\bibfnamefont {L.}~\bibnamefont
  {Bergstrom}},\ }\href {\doibase 10.1016/0370-2693(89)90585-6} {\bibfield
  {journal} {\bibinfo  {journal} {Phys.Lett.}\ }\textbf {\bibinfo {volume}
  {B225}},\ \bibinfo {pages} {372} (\bibinfo {year} {1989})}\BibitemShut
  {NoStop}%
\bibitem [{\citenamefont {Bringmann}\ \emph {et~al.}(2008)\citenamefont
  {Bringmann}, \citenamefont {Bergstrom},\ and\ \citenamefont
  {Edsjo}}]{Bringmann:2007nk}%
  \BibitemOpen
  \bibfield  {author} {\bibinfo {author} {\bibfnamefont {T.}~\bibnamefont
  {Bringmann}}, \bibinfo {author} {\bibfnamefont {L.}~\bibnamefont
  {Bergstrom}}, \ and\ \bibinfo {author} {\bibfnamefont {J.}~\bibnamefont
  {Edsjo}},\ }\href {\doibase 10.1088/1126-6708/2008/01/049} {\bibfield
  {journal} {\bibinfo  {journal} {JHEP}\ }\textbf {\bibinfo {volume} {0801}},\
  \bibinfo {pages} {049} (\bibinfo {year} {2008})},\ \Eprint
  {http://arxiv.org/abs/0710.3169} {arXiv:0710.3169 [hep-ph]} \BibitemShut
  {NoStop}%
\bibitem [{\citenamefont {Ciafaloni}\ \emph
  {et~al.}(2011{\natexlab{a}})\citenamefont {Ciafaloni}, \citenamefont
  {Comelli}, \citenamefont {Riotto}, \citenamefont {Sala}, \citenamefont
  {Strumia} \emph {et~al.}}]{Ciafaloni:2010ti}%
  \BibitemOpen
  \bibfield  {author} {\bibinfo {author} {\bibfnamefont {P.}~\bibnamefont
  {Ciafaloni}}, \bibinfo {author} {\bibfnamefont {D.}~\bibnamefont {Comelli}},
  \bibinfo {author} {\bibfnamefont {A.}~\bibnamefont {Riotto}}, \bibinfo
  {author} {\bibfnamefont {F.}~\bibnamefont {Sala}}, \bibinfo {author}
  {\bibfnamefont {A.}~\bibnamefont {Strumia}},  \emph {et~al.},\ }\href
  {\doibase 10.1088/1475-7516/2011/03/019} {\bibfield  {journal} {\bibinfo
  {journal} {JCAP}\ }\textbf {\bibinfo {volume} {1103}},\ \bibinfo {pages}
  {019} (\bibinfo {year} {2011}{\natexlab{a}})},\ \Eprint
  {http://arxiv.org/abs/1009.0224} {arXiv:1009.0224 [hep-ph]} \BibitemShut
  {NoStop}%
\bibitem [{\citenamefont {Ciafaloni}\ \emph
  {et~al.}(2011{\natexlab{b}})\citenamefont {Ciafaloni}, \citenamefont
  {Cirelli}, \citenamefont {Comelli}, \citenamefont {De~Simone}, \citenamefont
  {Riotto} \emph {et~al.}}]{Ciafaloni:2011sa}%
  \BibitemOpen
  \bibfield  {author} {\bibinfo {author} {\bibfnamefont {P.}~\bibnamefont
  {Ciafaloni}}, \bibinfo {author} {\bibfnamefont {M.}~\bibnamefont {Cirelli}},
  \bibinfo {author} {\bibfnamefont {D.}~\bibnamefont {Comelli}}, \bibinfo
  {author} {\bibfnamefont {A.}~\bibnamefont {De~Simone}}, \bibinfo {author}
  {\bibfnamefont {A.}~\bibnamefont {Riotto}},  \emph {et~al.},\ }\href
  {\doibase 10.1088/1475-7516/2011/06/018} {\bibfield  {journal} {\bibinfo
  {journal} {JCAP}\ }\textbf {\bibinfo {volume} {1106}},\ \bibinfo {pages}
  {018} (\bibinfo {year} {2011}{\natexlab{b}})},\ \Eprint
  {http://arxiv.org/abs/1104.2996} {arXiv:1104.2996 [hep-ph]} \BibitemShut
  {NoStop}%
\bibitem [{\citenamefont {Arkani-Hamed}\ \emph {et~al.}(2005)\citenamefont
  {Arkani-Hamed}, \citenamefont {Dimopoulos},\ and\ \citenamefont
  {Kachru}}]{ArkaniHamed:2005yv}%
  \BibitemOpen
  \bibfield  {author} {\bibinfo {author} {\bibfnamefont {N.}~\bibnamefont
  {Arkani-Hamed}}, \bibinfo {author} {\bibfnamefont {S.}~\bibnamefont
  {Dimopoulos}}, \ and\ \bibinfo {author} {\bibfnamefont {S.}~\bibnamefont
  {Kachru}},\ }\href@noop {} {\  (\bibinfo {year} {2005})},\ \Eprint
  {http://arxiv.org/abs/hep-th/0501082} {arXiv:hep-th/0501082} \BibitemShut
  {NoStop}%
\bibitem [{\citenamefont {Mahbubani}\ and\ \citenamefont
  {Senatore}(2006)}]{Mahbubani:2005pt}%
  \BibitemOpen
  \bibfield  {author} {\bibinfo {author} {\bibfnamefont {R.}~\bibnamefont
  {Mahbubani}}\ and\ \bibinfo {author} {\bibfnamefont {L.}~\bibnamefont
  {Senatore}},\ }\href {\doibase 10.1103/PhysRevD.73.043510} {\bibfield
  {journal} {\bibinfo  {journal} {Phys.Rev.}\ }\textbf {\bibinfo {volume}
  {D73}},\ \bibinfo {pages} {043510} (\bibinfo {year} {2006})},\ \Eprint
  {http://arxiv.org/abs/hep-ph/0510064} {arXiv:hep-ph/0510064 [hep-ph]}
  \BibitemShut {NoStop}%
\bibitem [{\citenamefont {Enberg}\ \emph {et~al.}(2007)\citenamefont {Enberg},
  \citenamefont {Fox}, \citenamefont {Hall}, \citenamefont {Papaioannou},\ and\
  \citenamefont {Papucci}}]{Enberg:2007rp}%
  \BibitemOpen
  \bibfield  {author} {\bibinfo {author} {\bibfnamefont {R.}~\bibnamefont
  {Enberg}}, \bibinfo {author} {\bibfnamefont {P.}~\bibnamefont {Fox}},
  \bibinfo {author} {\bibfnamefont {L.}~\bibnamefont {Hall}}, \bibinfo {author}
  {\bibfnamefont {A.}~\bibnamefont {Papaioannou}}, \ and\ \bibinfo {author}
  {\bibfnamefont {M.}~\bibnamefont {Papucci}},\ }\href {\doibase
  10.1088/1126-6708/2007/11/014} {\bibfield  {journal} {\bibinfo  {journal}
  {JHEP}\ }\textbf {\bibinfo {volume} {0711}},\ \bibinfo {pages} {014}
  (\bibinfo {year} {2007})},\ \Eprint {http://arxiv.org/abs/0706.0918}
  {arXiv:0706.0918 [hep-ph]} \BibitemShut {NoStop}%
\bibitem [{\citenamefont {D'Eramo}(2007)}]{D'Eramo:2007ga}%
  \BibitemOpen
  \bibfield  {author} {\bibinfo {author} {\bibfnamefont {F.}~\bibnamefont
  {D'Eramo}},\ }\href {\doibase 10.1103/PhysRevD.76.083522} {\bibfield
  {journal} {\bibinfo  {journal} {Phys.Rev.}\ }\textbf {\bibinfo {volume}
  {D76}},\ \bibinfo {pages} {083522} (\bibinfo {year} {2007})},\ \Eprint
  {http://arxiv.org/abs/0705.4493} {arXiv:0705.4493 [hep-ph]} \BibitemShut
  {NoStop}%
\bibitem [{\citenamefont {Cohen}\ \emph {et~al.}(2011)\citenamefont {Cohen},
  \citenamefont {Kearney}, \citenamefont {Pierce},\ and\ \citenamefont
  {Tucker-Smith}}]{Cohen:2011ec}%
  \BibitemOpen
  \bibfield  {author} {\bibinfo {author} {\bibfnamefont {T.}~\bibnamefont
  {Cohen}}, \bibinfo {author} {\bibfnamefont {J.}~\bibnamefont {Kearney}},
  \bibinfo {author} {\bibfnamefont {A.}~\bibnamefont {Pierce}}, \ and\ \bibinfo
  {author} {\bibfnamefont {D.}~\bibnamefont {Tucker-Smith}},\ }\href@noop {} {\
   (\bibinfo {year} {2011})},\ \Eprint {http://arxiv.org/abs/1109.2604}
  {arXiv:1109.2604 [hep-ph]} \BibitemShut {NoStop}%
\bibitem [{\citenamefont {Fayet}(1974)}]{Fayet:1974fj}%
  \BibitemOpen
  \bibfield  {author} {\bibinfo {author} {\bibfnamefont {P.}~\bibnamefont
  {Fayet}},\ }\href {\doibase 10.1016/0550-3213(74)90113-8} {\bibfield
  {journal} {\bibinfo  {journal} {Nucl. Phys.}\ }\textbf {\bibinfo {volume}
  {B78}},\ \bibinfo {pages} {14} (\bibinfo {year} {1974})}\BibitemShut
  {NoStop}%
\bibitem [{\citenamefont {Alwall}\ \emph {et~al.}(2007)\citenamefont {Alwall},
  \citenamefont {Demin}, \citenamefont {de~Visscher}, \citenamefont {Frederix},
  \citenamefont {Herquet} \emph {et~al.}}]{Alwall:2007st}%
  \BibitemOpen
  \bibfield  {author} {\bibinfo {author} {\bibfnamefont {J.}~\bibnamefont
  {Alwall}}, \bibinfo {author} {\bibfnamefont {P.}~\bibnamefont {Demin}},
  \bibinfo {author} {\bibfnamefont {S.}~\bibnamefont {de~Visscher}}, \bibinfo
  {author} {\bibfnamefont {R.}~\bibnamefont {Frederix}}, \bibinfo {author}
  {\bibfnamefont {M.}~\bibnamefont {Herquet}},  \emph {et~al.},\ }\href
  {\doibase 10.1088/1126-6708/2007/09/028} {\bibfield  {journal} {\bibinfo
  {journal} {JHEP}\ }\textbf {\bibinfo {volume} {0709}},\ \bibinfo {pages}
  {028} (\bibinfo {year} {2007})},\ \Eprint {http://arxiv.org/abs/0706.2334}
  {arXiv:0706.2334 [hep-ph]} \BibitemShut {NoStop}%
\bibitem [{\citenamefont {Sjostrand}\ \emph {et~al.}(2006)\citenamefont
  {Sjostrand}, \citenamefont {Mrenna},\ and\ \citenamefont
  {Skands}}]{Sjostrand:2006za}%
  \BibitemOpen
  \bibfield  {author} {\bibinfo {author} {\bibfnamefont {T.}~\bibnamefont
  {Sjostrand}}, \bibinfo {author} {\bibfnamefont {S.}~\bibnamefont {Mrenna}}, \
  and\ \bibinfo {author} {\bibfnamefont {P.~Z.}\ \bibnamefont {Skands}},\
  }\href {\doibase 10.1088/1126-6708/2006/05/026} {\bibfield  {journal}
  {\bibinfo  {journal} {JHEP}\ }\textbf {\bibinfo {volume} {0605}},\ \bibinfo
  {pages} {026} (\bibinfo {year} {2006})},\ \Eprint
  {http://arxiv.org/abs/hep-ph/0603175} {arXiv:hep-ph/0603175 [hep-ph]}
  \BibitemShut {NoStop}%
\bibitem [{\citenamefont {Edsjo}()}]{wimpsimcode}%
  \BibitemOpen
  \bibfield  {author} {\bibinfo {author} {\bibfnamefont {J.}~\bibnamefont
  {Edsjo}},\ }\href {http://www.physto.se/~edsjo/wimpsim/} {\bibinfo  {journal}
  {WimpSim Neutrino Monte Carlo, Nusigma Neutrino Monte Carlo}\ }\BibitemShut
  {NoStop}%
\bibitem [{\citenamefont {Maltoni}\ \emph {et~al.}(2004)\citenamefont
  {Maltoni}, \citenamefont {Schwetz}, \citenamefont {Tortola},\ and\
  \citenamefont {Valle}}]{Maltoni:2004ei}%
  \BibitemOpen
\bibfield  {journal} {  }\bibfield  {author} {\bibinfo {author} {\bibfnamefont
  {M.}~\bibnamefont {Maltoni}}, \bibinfo {author} {\bibfnamefont
  {T.}~\bibnamefont {Schwetz}}, \bibinfo {author} {\bibfnamefont
  {M.}~\bibnamefont {Tortola}}, \ and\ \bibinfo {author} {\bibfnamefont
  {J.}~\bibnamefont {Valle}},\ }\href {\doibase 10.1088/1367-2630/6/1/122}
  {\bibfield  {journal} {\bibinfo  {journal} {New J.Phys.}\ }\textbf {\bibinfo
  {volume} {6}},\ \bibinfo {pages} {122} (\bibinfo {year} {2004})},\ \Eprint
  {http://arxiv.org/abs/hep-ph/0405172} {arXiv:hep-ph/0405172 [hep-ph]}
  \BibitemShut {NoStop}%
\bibitem [{\citenamefont {Dreiner}\ \emph {et~al.}(2010)\citenamefont
  {Dreiner}, \citenamefont {Haber},\ and\ \citenamefont
  {Martin}}]{Dreiner:2008tw}%
  \BibitemOpen
  \bibfield  {author} {\bibinfo {author} {\bibfnamefont {H.~K.}\ \bibnamefont
  {Dreiner}}, \bibinfo {author} {\bibfnamefont {H.~E.}\ \bibnamefont {Haber}},
  \ and\ \bibinfo {author} {\bibfnamefont {S.~P.}\ \bibnamefont {Martin}},\
  }\href {\doibase 10.1016/j.physrep.2010.05.002} {\bibfield  {journal}
  {\bibinfo  {journal} {Phys.Rept.}\ }\textbf {\bibinfo {volume} {494}},\
  \bibinfo {pages} {1} (\bibinfo {year} {2010})},\ \Eprint
  {http://arxiv.org/abs/0812.1594} {arXiv:0812.1594 [hep-ph]} \BibitemShut
  {NoStop}%
\bibitem [{\citenamefont {Barger}\ \emph {et~al.}(2011)\citenamefont {Barger},
  \citenamefont {Gao},\ and\ \citenamefont {Marfatia}}]{Barger:2011em}%
  \BibitemOpen
  \bibfield  {author} {\bibinfo {author} {\bibfnamefont {V.}~\bibnamefont
  {Barger}}, \bibinfo {author} {\bibfnamefont {Y.}~\bibnamefont {Gao}}, \ and\
  \bibinfo {author} {\bibfnamefont {D.}~\bibnamefont {Marfatia}},\ }\href
  {\doibase 10.1103/PhysRevD.83.055012} {\bibfield  {journal} {\bibinfo
  {journal} {Phys.Rev.}\ }\textbf {\bibinfo {volume} {D83}},\ \bibinfo {pages}
  {055012} (\bibinfo {year} {2011})},\ \Eprint {http://arxiv.org/abs/1101.4410}
  {arXiv:1101.4410 [hep-ph]} \BibitemShut {NoStop}%
\bibitem [{\citenamefont {Belanger}\ \emph {et~al.}(2011)\citenamefont
  {Belanger}, \citenamefont {Boudjema}, \citenamefont {Brun}, \citenamefont
  {Pukhov}, \citenamefont {Rosier-Lees} \emph {et~al.}}]{Belanger:2010gh}%
  \BibitemOpen
  \bibfield  {author} {\bibinfo {author} {\bibfnamefont {G.}~\bibnamefont
  {Belanger}}, \bibinfo {author} {\bibfnamefont {F.}~\bibnamefont {Boudjema}},
  \bibinfo {author} {\bibfnamefont {P.}~\bibnamefont {Brun}}, \bibinfo {author}
  {\bibfnamefont {A.}~\bibnamefont {Pukhov}}, \bibinfo {author} {\bibfnamefont
  {S.}~\bibnamefont {Rosier-Lees}},  \emph {et~al.},\ }\href {\doibase
  10.1016/j.cpc.2010.11.033} {\bibfield  {journal} {\bibinfo  {journal}
  {Comput.Phys.Commun.}\ }\textbf {\bibinfo {volume} {182}},\ \bibinfo {pages}
  {842} (\bibinfo {year} {2011})},\ \Eprint {http://arxiv.org/abs/1004.1092}
  {arXiv:1004.1092 [hep-ph]} \BibitemShut {NoStop}%
\bibitem [{\citenamefont {Jarosik}\ \emph {et~al.}(2011)\citenamefont {Jarosik}
  \emph {et~al.}}]{Jarosik:2010iu}%
  \BibitemOpen
  \bibfield  {author} {\bibinfo {author} {\bibfnamefont {N.}~\bibnamefont
  {Jarosik}} \emph {et~al.},\ }\href {\doibase 10.1088/0067-0049/192/2/14}
  {\bibfield  {journal} {\bibinfo  {journal} {Astrophys. J. Suppl.}\ }\textbf
  {\bibinfo {volume} {192}},\ \bibinfo {pages} {14} (\bibinfo {year} {2011})},\
  \Eprint {http://arxiv.org/abs/1001.4744} {arXiv:1001.4744 [astro-ph.CO]}
  \BibitemShut {NoStop}%
\bibitem [{\citenamefont {Tanaka}\ \emph {et~al.}(2011)\citenamefont {Tanaka}
  \emph {et~al.}}]{Tanaka:2011uf}%
  \BibitemOpen
  \bibfield  {author} {\bibinfo {author} {\bibfnamefont {T.}~\bibnamefont
  {Tanaka}} \emph {et~al.} (\bibinfo {collaboration} {Super-Kamiokande
  Collaboration}),\ }\href {\doibase 10.1088/0004-637X/742/2/78} {\bibfield
  {journal} {\bibinfo  {journal} {Astrophys.J.}\ }\textbf {\bibinfo {volume}
  {742}},\ \bibinfo {pages} {78} (\bibinfo {year} {2011})},\ \bibinfo {note}
  {long author list - awaiting processing},\ \Eprint
  {http://arxiv.org/abs/1108.3384} {arXiv:1108.3384 [astro-ph.HE]} \BibitemShut
  {NoStop}%
\bibitem [{\citenamefont {Weber}\ and\ \citenamefont
  {de~Boer}(2010)}]{Weber:2009pt}%
  \BibitemOpen
  \bibfield  {author} {\bibinfo {author} {\bibfnamefont {M.}~\bibnamefont
  {Weber}}\ and\ \bibinfo {author} {\bibfnamefont {W.}~\bibnamefont
  {de~Boer}},\ }\href {\doibase 10.1051/0004-6361/200913381} {\bibfield
  {journal} {\bibinfo  {journal} {Astron.Astrophys.}\ }\textbf {\bibinfo
  {volume} {509}},\ \bibinfo {pages} {A25} (\bibinfo {year} {2010})},\ \Eprint
  {http://arxiv.org/abs/0910.4272} {arXiv:0910.4272 [astro-ph.CO]} \BibitemShut
  {NoStop}%
\bibitem [{\citenamefont {Catena}\ and\ \citenamefont
  {Ullio}(2010)}]{Catena:2009mf}%
  \BibitemOpen
  \bibfield  {author} {\bibinfo {author} {\bibfnamefont {R.}~\bibnamefont
  {Catena}}\ and\ \bibinfo {author} {\bibfnamefont {P.}~\bibnamefont {Ullio}},\
  }\href {\doibase 10.1088/1475-7516/2010/08/004} {\bibfield  {journal}
  {\bibinfo  {journal} {JCAP}\ }\textbf {\bibinfo {volume} {1008}},\ \bibinfo
  {pages} {004} (\bibinfo {year} {2010})},\ \Eprint
  {http://arxiv.org/abs/0907.0018} {arXiv:0907.0018 [astro-ph.CO]} \BibitemShut
  {NoStop}%
\bibitem [{\citenamefont {Widrow}\ \emph {et~al.}(2008)\citenamefont {Widrow},
  \citenamefont {Pym},\ and\ \citenamefont {Dubinski}}]{Widrow:2008yg}%
  \BibitemOpen
  \bibfield  {author} {\bibinfo {author} {\bibfnamefont {L.~M.}\ \bibnamefont
  {Widrow}}, \bibinfo {author} {\bibfnamefont {B.}~\bibnamefont {Pym}}, \ and\
  \bibinfo {author} {\bibfnamefont {J.}~\bibnamefont {Dubinski}},\ }\href
  {\doibase 10.1086/587636} {\bibfield  {journal} {\bibinfo  {journal}
  {Astrophys.J.}\ }\textbf {\bibinfo {volume} {679}},\ \bibinfo {pages} {1239}
  (\bibinfo {year} {2008})},\ \Eprint {http://arxiv.org/abs/0801.3414}
  {arXiv:0801.3414 [astro-ph]} \BibitemShut {NoStop}%
\bibitem [{\citenamefont {McMillan}\ and\ \citenamefont
  {Binney}(2009)}]{McMillan:2009yr}%
  \BibitemOpen
  \bibfield  {author} {\bibinfo {author} {\bibfnamefont {P.~J.}\ \bibnamefont
  {McMillan}}\ and\ \bibinfo {author} {\bibfnamefont {J.~J.}\ \bibnamefont
  {Binney}},\ }\href@noop {} {\  (\bibinfo {year} {2009})},\ \Eprint
  {http://arxiv.org/abs/0907.4685} {arXiv:0907.4685 [astro-ph.GA]} \BibitemShut
  {NoStop}%
\bibitem [{\citenamefont {Bovy}\ \emph {et~al.}(2009)\citenamefont {Bovy},
  \citenamefont {Hogg},\ and\ \citenamefont {Rix}}]{Bovy:2009dr}%
  \BibitemOpen
  \bibfield  {author} {\bibinfo {author} {\bibfnamefont {J.}~\bibnamefont
  {Bovy}}, \bibinfo {author} {\bibfnamefont {D.~W.}\ \bibnamefont {Hogg}}, \
  and\ \bibinfo {author} {\bibfnamefont {H.-W.}\ \bibnamefont {Rix}},\ }\href
  {\doibase 10.1088/0004-637X/704/2/1704} {\bibfield  {journal} {\bibinfo
  {journal} {Astrophys.J.}\ }\textbf {\bibinfo {volume} {704}},\ \bibinfo
  {pages} {1704} (\bibinfo {year} {2009})},\ \Eprint
  {http://arxiv.org/abs/0907.5423} {arXiv:0907.5423 [astro-ph.GA]} \BibitemShut
  {NoStop}%
\bibitem [{\citenamefont {Reid}\ \emph {et~al.}(2009)\citenamefont {Reid},
  \citenamefont {Menten}, \citenamefont {Zheng}, \citenamefont {Brunthaler},
  \citenamefont {Moscadelli} \emph {et~al.}}]{Reid:2009nj}%
  \BibitemOpen
  \bibfield  {author} {\bibinfo {author} {\bibfnamefont {M.}~\bibnamefont
  {Reid}}, \bibinfo {author} {\bibfnamefont {K.}~\bibnamefont {Menten}},
  \bibinfo {author} {\bibfnamefont {X.}~\bibnamefont {Zheng}}, \bibinfo
  {author} {\bibfnamefont {A.}~\bibnamefont {Brunthaler}}, \bibinfo {author}
  {\bibfnamefont {L.}~\bibnamefont {Moscadelli}},  \emph {et~al.},\ }\href
  {\doibase 10.1088/0004-637X/700/1/137} {\bibfield  {journal} {\bibinfo
  {journal} {Astrophys.J.}\ }\textbf {\bibinfo {volume} {700}},\ \bibinfo
  {pages} {137} (\bibinfo {year} {2009})},\ \Eprint
  {http://arxiv.org/abs/0902.3913} {arXiv:0902.3913 [astro-ph.GA]} \BibitemShut
  {NoStop}%
\bibitem [{\citenamefont {Green}(2011)}]{Green:2011bv}%
  \BibitemOpen
  \bibfield  {author} {\bibinfo {author} {\bibfnamefont {A.~M.}\ \bibnamefont
  {Green}},\ }\href@noop {} {\  (\bibinfo {year} {2011})},\ \Eprint
  {http://arxiv.org/abs/1112.0524} {arXiv:1112.0524 [astro-ph.CO]} \BibitemShut
  {NoStop}%
\bibitem [{\citenamefont {Yaguna}(2010)}]{Yaguna:2010hn}%
  \BibitemOpen
  \bibfield  {author} {\bibinfo {author} {\bibfnamefont {C.~E.}\ \bibnamefont
  {Yaguna}},\ }\href {\doibase 10.1103/PhysRevD.81.075024} {\bibfield
  {journal} {\bibinfo  {journal} {Phys.Rev.}\ }\textbf {\bibinfo {volume}
  {D81}},\ \bibinfo {pages} {075024} (\bibinfo {year} {2010})},\ \Eprint
  {http://arxiv.org/abs/1003.2730} {arXiv:1003.2730 [hep-ph]} \BibitemShut
  {NoStop}%
\bibitem [{\citenamefont {Felizardo}\ \emph {et~al.}(2011)\citenamefont
  {Felizardo}, \citenamefont {Girard}, \citenamefont {Morlat}, \citenamefont
  {Fernandes}, \citenamefont {Giuliani} \emph {et~al.}}]{Felizardo:2011uw}%
  \BibitemOpen
  \bibfield  {author} {\bibinfo {author} {\bibfnamefont {M.}~\bibnamefont
  {Felizardo}}, \bibinfo {author} {\bibfnamefont {T.}~\bibnamefont {Girard}},
  \bibinfo {author} {\bibfnamefont {T.}~\bibnamefont {Morlat}}, \bibinfo
  {author} {\bibfnamefont {A.}~\bibnamefont {Fernandes}}, \bibinfo {author}
  {\bibfnamefont {F.}~\bibnamefont {Giuliani}},  \emph {et~al.},\ }\href@noop
  {} {\  (\bibinfo {year} {2011})},\ \Eprint {http://arxiv.org/abs/1106.3014}
  {arXiv:1106.3014 [astro-ph.CO]} \BibitemShut {NoStop}%
\bibitem [{\citenamefont {Dahl}(2011)}]{COUPPproj}%
  \BibitemOpen
  \bibfield  {author} {\bibinfo {author} {\bibfnamefont {E.}~\bibnamefont
  {Dahl}},\ }\href@noop {} {\bibfield  {journal} {\bibinfo  {journal} {Talk at
  FNAL Joint Experimental-Theoretical Seminar,
  \url{http://theory.fnal.gov/jetp/talks/DAHL_COUPP_WineAndCheese2011.pdf}}\ }
  (\bibinfo {year} {2011})}\BibitemShut {NoStop}%
\bibitem [{ATL(2011)}]{ATLASmonojet}%
  \BibitemOpen
  \href@noop {} {\bibfield  {journal} {\bibinfo  {journal} {ATLAS
  Collaboration, ATLAS-CONF-2011-096,
  \url{http://cdsweb.cern.ch/record/1369187}}\ } (\bibinfo {year}
  {2011})}\BibitemShut {NoStop}%
\bibitem [{\citenamefont {Goodman}\ \emph {et~al.}(2010)\citenamefont
  {Goodman}, \citenamefont {Ibe}, \citenamefont {Rajaraman}, \citenamefont
  {Shepherd}, \citenamefont {Tait} \emph {et~al.}}]{Goodman:2010ku}%
  \BibitemOpen
  \bibfield  {author} {\bibinfo {author} {\bibfnamefont {J.}~\bibnamefont
  {Goodman}}, \bibinfo {author} {\bibfnamefont {M.}~\bibnamefont {Ibe}},
  \bibinfo {author} {\bibfnamefont {A.}~\bibnamefont {Rajaraman}}, \bibinfo
  {author} {\bibfnamefont {W.}~\bibnamefont {Shepherd}}, \bibinfo {author}
  {\bibfnamefont {T.~M.}\ \bibnamefont {Tait}},  \emph {et~al.},\ }\href
  {\doibase 10.1103/PhysRevD.82.116010} {\bibfield  {journal} {\bibinfo
  {journal} {Phys.Rev.}\ }\textbf {\bibinfo {volume} {D82}},\ \bibinfo {pages}
  {116010} (\bibinfo {year} {2010})},\ \Eprint {http://arxiv.org/abs/1008.1783}
  {arXiv:1008.1783 [hep-ph]} \BibitemShut {NoStop}%
\bibitem [{\citenamefont {Fox}\ \emph {et~al.}(2011)\citenamefont {Fox},
  \citenamefont {Harnik}, \citenamefont {Kopp},\ and\ \citenamefont
  {Tsai}}]{Fox:2011pm}%
  \BibitemOpen
  \bibfield  {author} {\bibinfo {author} {\bibfnamefont {P.~J.}\ \bibnamefont
  {Fox}}, \bibinfo {author} {\bibfnamefont {R.}~\bibnamefont {Harnik}},
  \bibinfo {author} {\bibfnamefont {J.}~\bibnamefont {Kopp}}, \ and\ \bibinfo
  {author} {\bibfnamefont {Y.}~\bibnamefont {Tsai}},\ }\href@noop {} {\
  (\bibinfo {year} {2011})},\ \Eprint {http://arxiv.org/abs/1109.4398}
  {arXiv:1109.4398 [hep-ph]} \BibitemShut {NoStop}%
\bibitem [{\citenamefont {Rajaraman}\ \emph {et~al.}(2011)\citenamefont
  {Rajaraman}, \citenamefont {Shepherd}, \citenamefont {Tait},\ and\
  \citenamefont {Wijangco}}]{Rajaraman:2011wf}%
  \BibitemOpen
  \bibfield  {author} {\bibinfo {author} {\bibfnamefont {A.}~\bibnamefont
  {Rajaraman}}, \bibinfo {author} {\bibfnamefont {W.}~\bibnamefont {Shepherd}},
  \bibinfo {author} {\bibfnamefont {T.~M.}\ \bibnamefont {Tait}}, \ and\
  \bibinfo {author} {\bibfnamefont {A.~M.}\ \bibnamefont {Wijangco}},\
  }\href@noop {} {\  (\bibinfo {year} {2011})},\ \Eprint
  {http://arxiv.org/abs/1108.1196} {arXiv:1108.1196 [hep-ph]} \BibitemShut
  {NoStop}%
\bibitem [{\citenamefont {Goodman}\ and\ \citenamefont
  {Shepherd}(2011)}]{Goodman:2011jq}%
  \BibitemOpen
  \bibfield  {author} {\bibinfo {author} {\bibfnamefont {J.}~\bibnamefont
  {Goodman}}\ and\ \bibinfo {author} {\bibfnamefont {W.}~\bibnamefont
  {Shepherd}},\ }\href@noop {} {\  (\bibinfo {year} {2011})},\ \Eprint
  {http://arxiv.org/abs/1111.2359} {arXiv:1111.2359 [hep-ph]} \BibitemShut
  {NoStop}%
\bibitem [{\citenamefont {Desai}\ \emph {et~al.}(2004)\citenamefont {Desai}
  \emph {et~al.}}]{Desai:2004pq}%
  \BibitemOpen
  \bibfield  {author} {\bibinfo {author} {\bibfnamefont {S.}~\bibnamefont
  {Desai}} \emph {et~al.} (\bibinfo {collaboration} {Super-Kamiokande
  Collaboration}),\ }\href {\doibase 10.1103/PhysRevD.70.083523,
  10.1103/PhysRevD.70.109901, 10.1103/PhysRevD.70.083523,
  10.1103/PhysRevD.70.109901} {\bibfield  {journal} {\bibinfo  {journal}
  {Phys.Rev.}\ }\textbf {\bibinfo {volume} {D70}},\ \bibinfo {pages} {083523}
  (\bibinfo {year} {2004})},\ \Eprint {http://arxiv.org/abs/hep-ex/0404025}
  {arXiv:hep-ex/0404025 [hep-ex]} \BibitemShut {NoStop}%
\bibitem [{\citenamefont {Aprile}\ \emph {et~al.}(2009)\citenamefont {Aprile},
  \citenamefont {Baudis}, \citenamefont {Choi}, \citenamefont {Giboni},
  \citenamefont {Lim} \emph {et~al.}}]{Aprile:2008rc}%
  \BibitemOpen
  \bibfield  {author} {\bibinfo {author} {\bibfnamefont {E.}~\bibnamefont
  {Aprile}}, \bibinfo {author} {\bibfnamefont {L.}~\bibnamefont {Baudis}},
  \bibinfo {author} {\bibfnamefont {B.}~\bibnamefont {Choi}}, \bibinfo {author}
  {\bibfnamefont {K.}~\bibnamefont {Giboni}}, \bibinfo {author} {\bibfnamefont
  {K.}~\bibnamefont {Lim}},  \emph {et~al.},\ }\href {\doibase
  10.1103/PhysRevC.79.045807} {\bibfield  {journal} {\bibinfo  {journal}
  {Phys.Rev.}\ }\textbf {\bibinfo {volume} {C79}},\ \bibinfo {pages} {045807}
  (\bibinfo {year} {2009})},\ \Eprint {http://arxiv.org/abs/0810.0274}
  {arXiv:0810.0274 [astro-ph]} \BibitemShut {NoStop}%
\bibitem [{atl(2011)}]{atlashiggs}%
  \BibitemOpen
  \href@noop {} {\bibfield  {journal} {\bibinfo  {journal} {ATLAS
  Collaboration, ATLAS-CONF-2011-163,
  \url{http://cdsweb.cern.ch/record/1406358}}\ } (\bibinfo {year}
  {2011})}\BibitemShut {NoStop}%
\bibitem [{cms(2011)}]{cmshiggs}%
  \BibitemOpen
  \href@noop {} {\bibfield  {journal} {\bibinfo  {journal} {CMS Collaboration,
  CMS-PAS-HIG-11-032, \url{http://cdsweb.cern.ch/record/1406347?ln=en}}\ }
  (\bibinfo {year} {2011})}\BibitemShut {NoStop}%
\bibitem [{\citenamefont {Gondolo}\ and\ \citenamefont
  {Gelmini}(1991)}]{Gondolo:1990dk}%
  \BibitemOpen
  \bibfield  {author} {\bibinfo {author} {\bibfnamefont {P.}~\bibnamefont
  {Gondolo}}\ and\ \bibinfo {author} {\bibfnamefont {G.}~\bibnamefont
  {Gelmini}},\ }\href {\doibase 10.1016/0550-3213(91)90438-4} {\bibfield
  {journal} {\bibinfo  {journal} {Nucl.Phys.}\ }\textbf {\bibinfo {volume}
  {B360}},\ \bibinfo {pages} {145} (\bibinfo {year} {1991})}\BibitemShut
  {NoStop}%
\end{thebibliography}%
\end{document}